\documentclass[aps,prb,twocolumn,reprint,superscriptaddress,showkeys,floatfix,amsmath,amssymb]{revtex4-1}
\usepackage{epsfig}
\usepackage[utf8]{inputenc}
\usepackage{color,soul}
\usepackage{hyperref}
\usepackage{graphicx}
\usepackage{epstopdf}
\usepackage{hypernat}
\usepackage[normalem]{ulem}
\usepackage{xcolor}
\usepackage{float}
\usepackage{siunitx}
\usepackage[most]{tcolorbox}
\usepackage[export]{adjustbox}

\usepackage[
top=1in,
bottom=1in,
left=0.7in,
right=0.7in,
]{geometry}

\definecolor{darkgreen}{rgb}{0.0, 0.5, 0.0}
\definecolor{darkyellow}{RGB}{180,110,30}

\begin{document}

\preprint{APS/123-QED}

\title{Frustration-Stabilized Hexamerons and Their Topological Magnons in Janus MXene}

\author{C. B. Subba} 
\affiliation{Advanced Functional Materials \& Simulation Lab (AFMSL), Department of Physics, Mizoram University, Aizawl-796004, India}
\author{Z. Pachuau}%
\affiliation{Department of Physics, Mizoram University, Aizawl-796004, India.
}%

\author{Samy Brahimi}
 \affiliation{Laboratoire de Physique et Chimie Quantique, Universite Mouloud Mammeri de Tizi-Ouzou, 15000 Tizi-Ouzou, Algeria}

 \author{Samir Lounis}
\email[Samir Lounis:]{samir.lounis@physik.uni-halle.de}
\affiliation{Institute of Physics and Halle-Berlin-Regensburg Cluster of Excellence CCE, Martin-Luther-University Halle-Wittenberg, 06099 Halle (Saale), Germany}

\author{D. P. Rai}
\email [D. P. Rai:]{dibyaprakashrai@gmail.com}	
\affiliation{Advanced Functional Materials \& Simulation Lab (AFMSL), Department of Physics, Mizoram University, Aizawl-796004, India}

\begin{abstract}
Frustrated magnets on triangular lattices favour noncollinear order, and when inversion symmetry is broken they can host localized topological textures on top of that order. Here we show that the Janus MXene MnHf$_2$C$_2$FI is such a system. Combining first-principles calculations with atomistic spin simulations, we find that magnetism is confined to the Mn layer, where a strong antiferromagnetic nearest-neighbor exchange together with a sizable Dzyaloshinskii--Moriya interaction stabilizes a frustrated $120^\circ$ N\'eel ground state. On top of this ground state we identify a metastable hexameron: a composite of six merons and antimerons, one vortex--antivortex pair on each of the three sublattices, whose sublattice-resolved topological charges compensate. Linear spin-wave theory confirms that the hexameron is a local energy minimum, and its noncoplanar cores act as potential wells that bind discrete in-gap magnon modes below the bulk continuum. In a periodic hexameron array the magnon bands carry nonzero Chern numbers, and the resulting Berry curvature produces a magnon thermal Hall response whose sign depends on which of the two time-reversal partners of the hexameron is realized. Our results establish Janus MXenes as a two-dimensional platform in which frustration, broken inversion symmetry, and spin--orbit coupling combine to produce metastable antiferromagnetic solitons with distinct spectroscopic and transport fingerprints.

\end{abstract}
\maketitle


\section{INTRODUCTION} 
Magnetic frustration plays a crucial role in the emergence of exotic magnetic phenomena, including noncoplanar magnetic order\cite{abuawwad2022noncollinear,abuawwad2024electrical,chaix2026triangular,okamura2026giant}, quantum spin liquids~\cite{ramirez1994strongly,balents2010spin,buessen2018quantum,diep2013frustrated,laforge2013quasispin,chatterjee2026semi,jena2026spin,li2026kitaev}, topological spin freezing and spin-glass behavior~\cite{jena2026topological}, antiferromagnetic (AFM) topological textures\cite{nagaosa2013topological,nayak2017magnetic,yuan2020intrinsic,yuan2020intrinsic,amin2023antiferromagnetic,bhukta2024homochiral,zhou2025triple,souza2026dynamics,li2026hybrid} and incommensurate magnetic order~\cite{chung2005statics,hromov2025incommensurate,zhou2025triple,cheng2026interwoven,ma2026coexistence}. Frustration arises when competing interactions or the geometry of the lattice prevent all magnetic interactions from being simultaneously satisfied. These situations are commonly referred to as interaction-induced frustration and geometric frustration, respectively. One of the consequences of magnetic frustration is the prevention of the Néel state or magnetic states characterized by long-range order. Consequently, one expects the magnetic susceptibility to exhibit paramagnetic behavior, persisting even down to the lowest temperatures. This implies that the system fails to develop long-range magnetic correlations. Frustrated systems (such as spin liquids) differ from standard paramagnetic states in which magnetic moments fluctuate independently, by the presence of short-range correlations. This led Villain \cite{villain1979insulating} to introduce the concept of "cooperative paramagnetism.". A prototypical example is the nearest-neighbor Ising antiferromagnet on a triangular lattice, where the three spins on each elementary triangle cannot satisfy all three antiferromagnetic interactions simultaneously. As a result, the system has an extensively degenerate manifold of ground states and a finite residual entropy in the thermodynamic limit. Wannier's exact solution, later corrected by an erratum, gives a residual entropy of \(S_0/N \simeq 0.3231\,k_{\mathrm B}\) per spin~\cite{PhysRev.79.357,wannier1973antiferromagnetism}, about \(46.6\%\) of the infinite-temperature entropy \(k_{\mathrm B}\ln 2\) per spin. For this reason, the two-dimensional triangular lattice is the prototypical frustrated geometry, especially when it hosts antiferromagnetically interacting spin moments $1/2$~\cite{misguich1999spin}.
The interplay of frustration, competing exchange interactions, and spin–orbit coupling (SOC) can lead to a wider variety of magnetic states, including noncollinear and noncoplanar order as well as topologically nontrivial spin textures\cite{aldarawsheh2023intrinsic,zhou2025triple}. In triangular and kagome magnets, competing exchange interactions can still favor noncollinear configurations even in the absence of SOC\cite{starykh2015unusual,zhou2024kagomerization,wang2025intriguing,barik2026magnetic}. When SOC is present, the Dzyaloshinskii–Moriya interaction (DMI), which originates from the antisymmetric part of the exchange interaction and is allowed when the relevant bond lacks inversion symmetry, can favor a particular sense of spin rotation. In competition with symmetric exchange, magnetic anisotropy, and external magnetic fields, DMI can stabilize chiral spin textures, including magnetic skyrmions\cite{heinze2011spontaneous,seki2012formation,nagaosa2013topological,birch2022history,kameda2026interactions}. In other words, inversion symmetry breaking and spin-orbit coupling work together to create a spin texture in momentum space. These considerations motivate studying antiferromagnetic (AFM) frustrated magnetic systems with broken inversion symmetry as potential hosts of unconventional magnetic order and topological spin textures.

Recent progress in the experimental study on topological magnetism~\cite{jungwirth2016antiferromagnetic,gomonay2017concepts,nvemec2018antiferromagnetic,gomonay2018antiferromagnetic,aldarawsheh2022emergence}, driven by its potential for information technology, has renewed interest in antiferromagnets (AFMs)\cite{nunez2006theory,gomonay2012symmetry,zhang2016antiferromagnetic,jungwirth2016antiferromagnetic,dal2024antiferromagnetic}. Their spin dynamics can be several orders of magnitude faster than those of ferromagnets (FMs)~\cite{kampfrath2011coherent,baierl2016terahertz,bhattacharjee2018neel}. In particular, AFM spin textures with swirling, topologically nontrivial configurations are attractive because they can behave as particle-like objects and may serve as robust information carriers. Manipulating them is expected to require low energy~\cite{rosales2015three,barker2016static,zhang2016antiferromagnetic,jin2016dynamics,gomonay2018antiferromagnetic,fernandes2019skyrmions,legrand2020room,shen2020current,juge2022skyrmions,silva2021antiferromagnetic}. Moreover, AFM textures can be driven by electric currents~\cite{barker2016static,akosa2017theory,zhang2016antiferromagnetic} without the transverse deflection associated with the skyrmion Hall effect, which affects their FM counterparts~\cite{lin2013particle,litzius2017skyrmion}. Therefore, AFM skyrmions can travel long distances without reaching the sample edge. Consistent with these developments, researchers have experimentally realized synthetic AFM skyrmions in coupled magnetic bilayers, where two FM skyrmions in separate layers are antiferromagnetically coupled across a nonmagnetic spacer layer~\cite{dohi2019formation,chen2022controllable}. To date, intrinsic AFM skyrmions, i.e., skyrmions forming within a single AFM layer, have not been experimentally observed \cite{bhukta2026time,pan2024strain,li2025intrinsic}.

Since the discovery in 2011, two-dimensional transition-metal carbides and nitrides, known as MXenes~\cite{naguib2011two}, have been studied extensively. Their general formula is M$_{n+1}$X$_n$T$_x$, where M is an early transition metal, X is C or N, and T$_x$ denotes surface terminations such as \,--O, \,--OH, or \,--F. Their metallic conductivity, chemical stability, and tunable surface chemistry have made them attractive for energy-related applications~\cite{kumar2022methods,subba2025comprehensive,subba20252d,ajmal2025advancements}. Surface terminations strongly influence the band gap, Fermi level, and charge-transfer characteristics~\cite{nag2025modulation,MOKKATH2025112967,D5CP01402F,doi:10.1021/acs.jpclett.0c03710}, and heterostructure formation and defect engineering have improved light absorption and charge separation~\cite{subba20252d,prakash2025mxene,mahapatra2025artificial}. However, the magnetic properties of MXenes have received considerably less attention, even though MXenes containing magnetic transition metals such as Mn and Cr offer a natural platform for two-dimensional magnetism\cite{amrillah2023potential,garcia2023theoretical,limbu2025magnetic,kim2026magnetic}. Moreover, Janus MXenes can be synthesized\cite{akgenc2021tuning,chen2023two,ram2024tuning,helmi2025novel,qian2025biomimetic,lee2026asymmetrically,thasitha2025janus,faysal2026first}, in which asymmetric surface terminations break inversion symmetry, can enable spin--orbit-driven interactions such as the Dzyaloshinskii--Moriya interaction (DMI) and may host frustrated and noncollinear magnetic states~\cite{kabiraj2023realizing,mokkath2025field}. They therefore provide a platform for studying the interplay of frustrated exchange interactions, broken inversion symmetry, and spin--orbit coupling (SOC), which together promote unconventional magnetic order and topological spin textures\cite{aldarawsheh2023intrinsic,zhou2025triple}. 

To our knowledge, the magnetic textures that frustration can produce in Janus MXenes have not yet been explored systematically. Here we address this question for MnHf$_2$C$_2$FI, using atomistic spin simulations parameterized from first principles with a Hamiltonian that includes isotropic Heisenberg exchange, the DMI, uniaxial anisotropy, and the Zeeman interaction. The picture that emerges is simple. The Janus termination confines magnetism to the Mn layer; the dominant antiferromagnetic nearest-neighbor exchange turns this layer into a frustrated triangular antiferromagnet with a $120^\circ$ N\'eel ground state; and on top of this ground state a metastable intrinsic multimeronic texture, the hexameron, can be stabilized. The Néel state can be decomposed into three magnetic sublattices, each characterized by a FM magnetization oriented at 120° with respect to those of the other two sublattices. A hexameron is a composite magnetic texture in which an FM meronic pair forms on each sublattice, resulting in a total of six merons (or antimerons) embedded within a single AFM Néel meronic texture. This places MnHf$_2$C$_2$FI in the same family as the intrinsic N\'eel multimeronic textures previously predicted~\cite{aldarawsheh2023intrinsic,abuawwad2023crte}. We follow the hexameron as a function of magnetic field and of the microscopic parameters, and then turn to its excitations: the hexameron binds localized magnon modes, and a periodic array of hexamerons has topologically nontrivial magnon bands with a thermal Hall response that distinguishes the two time-reversal partners.

\section{COMPUTATIONAL DETAILS}
We performed first-principles calculations for the MXene using the Vienna Ab initio Simulation Package (VASP) \cite{kresse1996efficient,kresse1993ab} within the projector augmented-wave (PAW) method \cite{blochl1994projector}. The exchange-correlation functional was treated within the generalized gradient approximation of Perdew, Burke, and Ernzerhof (PBE) \cite{perdew1996generalized}. The localized Mn-$3d$ electrons were described using the rotationally invariant DFT+$U$ scheme of Dudarev et al. \cite{liechtenstein1995density,dudarev1998electron}, with $U_{\mathrm{eff}} = 3.9$ eV chosen based on previous reports~\cite{zhang2023giant,moore2024high,wang2006oxidation,jain2011formation}. A plane-wave cutoff of 500 eV was used, and the Brillouin zone was sampled with a $\Gamma$-centered $18\times18\times1$ $k$-mesh for the primitive cell. A vacuum layer of $\approx 24~\mathrm{\AA}$ was included to separate periodic images, and dipole corrections along the out-of-plane direction were applied to account for the intrinsic electric dipole of the Janus structure. Lattice parameters and atomic positions were relaxed until the total energy and residual forces were below $10^{-8}$ eV and 0.01 eV/\AA{}, respectively. Van der Waals corrections were included via DFT-D3 \cite{grimme2010consistent}.
To assess the dynamical stability of the Janus MnHf$_2$C$_2$FI monolayer, we computed the zone-centre ($\Gamma$-point) vibrational frequencies using density functional perturbation theory (DFPT)~\cite{PhysRevB.73.045112} as implemented in VASP. The structure was taken from a tightly relaxed structure, and the Brillouin zone was sampled with a $12 \times 12 \times 1$ $\Gamma$-centred Monkhorst--Pack grid for frequency calculation. Because the Janus structure breaks inversion symmetry, all symmetry operations were disabled ($\texttt{ISYM} = -1$) during the atomic displacements. A strict electronic convergence criterion of $\texttt{EDIFF} = 10^{-8}$~eV was used to obtain accurate interatomic force constants.

To extract the magnetic interaction parameters, we constructed a tight-binding Hamiltonian in the basis of maximally localized Wannier functions using the Wannier90 code~\cite{mostofi2008wannier90}. The exchange interactions were then evaluated with the Green's-function method within the magnetic force theorem~\cite{he2021tb2j,rezaei2026benchmarking}. We performed noncollinear calculations including spin–orbit coupling (SOC) for six constrained magnetization directions, denoted by \(a\), \(b\), \(c\), \(x\), \(y\), and \(z\) (see Fig.S10). The procedure used to reconstruct the full exchange tensor from these calculations is described in the Supplementary Material (see Fig.~S8). The resulting exchange tensors were decomposed into isotropic Heisenberg exchange, antisymmetric Dzyaloshinskii–Moriya interaction (DMI), and symmetric anisotropic exchange contributions, as described in Appendix \ref{atomistic_model} (Eq.\ref{J_decomposition}). The magnetic anisotropy energy (MAE) was evaluated from the total-energy difference between out-of-plane and in-plane magnetization configurations.

The magnetic configurations is described using a classical atomistic
spin Hamiltonian comprising isotropic Heisenberg exchange, uniaxial
magnetic anisotropy, the Zeeman interaction, and the
Dzyaloshinskii--Moriya interaction (DMI):
\begin{equation}
\begin{aligned}
\mathcal{H}_{\mathrm{spin}} ={}&
-\sum_{i<j} J_{ij}\,
\mathbf{S}_i\cdot\mathbf{S}_j
+K_u\sum_i(\mathbf{S}_i\cdot\hat{\mathbf{e}}_i)^2\\
&-\mu_S\sum_i\mathbf{S}_i\cdot\mathbf{B}_{\mathrm{app}}
-\sum_{i<j}\mathbf{D}_{ij}\cdot
(\mathbf{S}_i\times\mathbf{S}_j),
\end{aligned}
\label{eq:spin_hamiltonian}
\end{equation}
where $\mathbf{S}_i$ and $\mathbf{S}_j$ are unit vectors describing
the orientations of the atomic magnetic moments, $J_{ij}$ is the
isotropic exchange coupling, $K_u$ is the uniaxial anisotropy
constant, $\hat{\mathbf{e}}_i$ is the local easy-axis direction, a positive uniaxial magnetocrystalline anisotropy constant ($K > 0$) along the out-of-plane $[0,0,1]$ axis corresponds to an easy-plane anisotropy that penalizes out-of-plane spin alignment and favors moment rotation within the $xy$ surface plane,
$\mu_S$ is the atomic magnetic moment, $\mathbf{B}_{\mathrm{app}}$
is the applied magnetic field, and $\mathbf{D}_{ij}$ is the DMI
vector. The time evolution of the spin configurations was obtained
by integrating the Landau--Lifshitz--Gilbert (LLG) equation\cite{spiritCode,spiritPaper,evans2014atomistic},
\begin{equation}
\frac{d\mathbf{S}_i}{dt}
=-\frac{\gamma}{1+\alpha^2}
\left[
\mathbf{S}_i\times\mathbf{B}_i^{\mathrm{eff}}
+\alpha\,\mathbf{S}_i\times
\left(\mathbf{S}_i\times\mathbf{B}_i^{\mathrm{eff}}\right)
\right],
\label{eq:LLG}
\end{equation}
where $\gamma$ is the gyromagnetic ratio, $\alpha$ is the
dimensionless Gilbert damping parameter, and
$\mathbf{B}_i^{\mathrm{eff}}$ is the local effective magnetic field.
The LLG equation was integrated numerically using the Heun method to study the evolution of noncollinear magnetic configurations and topologically nontrivial spin textures. Further details on the exchange interaction and the construction of the spin Hamiltonian are provided in Appendix~\ref{atomistic_model}.

The magnetic parameters were then used to study the magnetic properties of the system through atomistic spin simulations using the Vampire package~\cite{evans2014atomistic} and the spirit code~\cite{spiritPaper,spiritCode}.

In VAMPIRE, temperature-dependent Monte Carlo simulations were performed using the DFT-relaxed $\mathrm{Hf_2MnC_2FI}$ supercell ($a = b = 3.33~\mathrm{\AA}$, $c = 32.83~\mathrm{\AA}$). The seven-atom-thick monolayer ($8.93~\mathrm{\AA}$) was separated from its periodic images by $23.89~\mathrm{\AA}$ of vacuum. A $15.0~\mathrm{nm} \times 15.0~\mathrm{nm} \times 3.28~\mathrm{nm}$ simulation box, corresponding to a single unit cell along $z$, was used with in-plane periodic boundary conditions and open boundaries along the $z$-axis. Spin moments were restricted to the single 2D Mn atomic plane to characterize the magnetic ordering and determine the transition temperature from the temperature dependence of the relevant magnetic observables.
The magnetic entropy was obtained by thermodynamic integration of the specific heat, as described in Appendix~\ref{Residual_Entropy}. We also examined the real-space spin configurations of the Janus MXene monolayer using spin dynamics governed by the Landau--Lifshitz--Gilbert (LLG) equation.
These large-scale spin dynamics simulations were performed using the \textsc{Spirit} code on a $1000 \times 1000 \times 1$ hexagonal grid ($10^6$ spins), corresponding to a lateral simulation area of approximately $332.8~\mathrm{nm} \times 288.2~\mathrm{nm}$ for the single-layer two-dimensional Mn monolayer ($a = 3.33~\mathrm{\AA}$).
The resulting multi-meron textures were characterized using the lattice-resolved topological charge density $Q_L$ and the corresponding integrated total topological charge.

The topological charge is defined as\cite{rybakov2019chiral,muller2019spirit}
\begin{equation}
  Q=\frac{1}{4\pi}\int \mathbf{n}\cdot\left(\partial_x\mathbf{n}\times
  \partial_y\mathbf{n}\right)\,dx\,dy,
  \label{eq:Q}
\end{equation}
where $\mathbf{n}=\mathbf{S}/|\mathbf{S}|$ is the unit vector of local 
magnetization. The case of a discrete atomistic lattice is discussed in the Supplementary Material.
The magnon spectra around the relaxed classical spin configurations were computed within the framework of linear spin-wave theory (LSWT) as implemented in the Magnopy package \cite{rybakov2026quantum}.

As a reference for how frustration shows up in thermodynamics, we first studied Ising models on triangular and honeycomb lattices. For Ising spins, thermodynamic integration gives the residual entropy, i.e., the ground-state degeneracy, directly. Detailed results and discussion are provided in the Supplementary Material (see sections II and III, Figs.~S1, S2, S5, and S6) and are compiled in Table I, where we evaluated the Curie temperature assuming purely ferromagnetic interactions in a system of size $1.6 \times 10^3$.

\section*{RESULTS AND DISCUSSION}
\subsection{Structural Properties}
\begin{figure}[H]
\includegraphics[width=\linewidth]{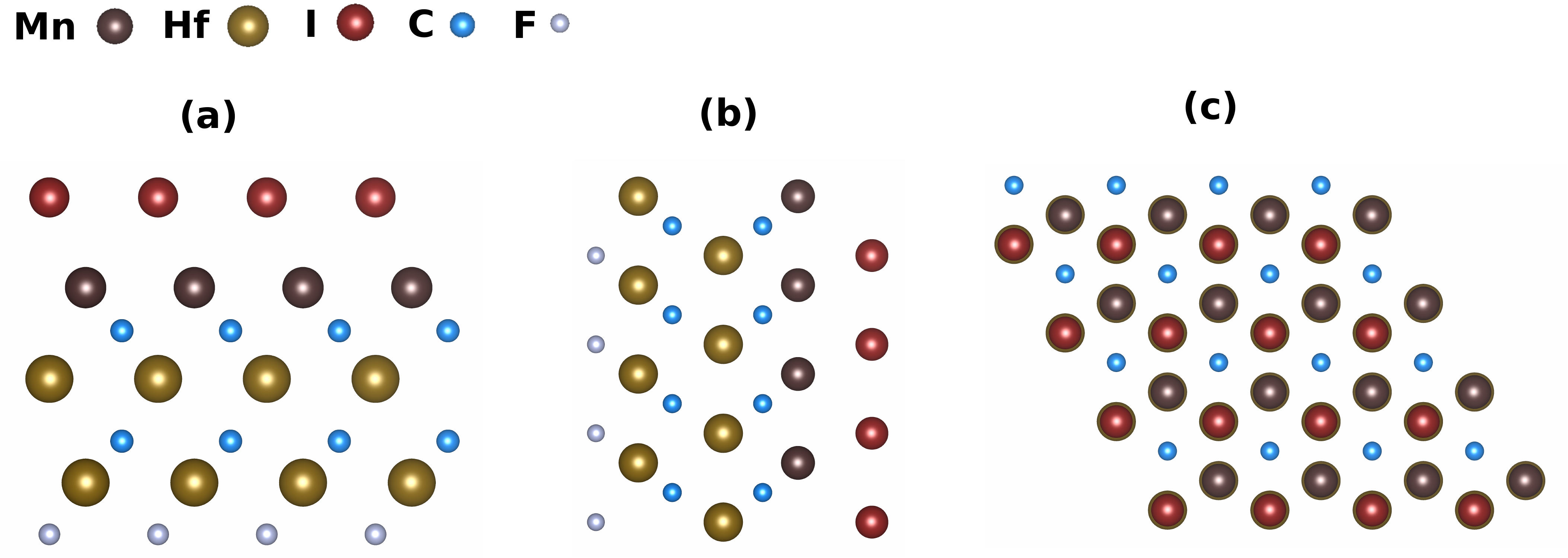}
\caption{Structure of the Janus MXene MnHf$_2$C$_2$FI. (a) Top view. (b,c) Side views along the $x$ and $y$ axes.}
\label{structure}
\end{figure}

\begin{figure*}
\includegraphics[width=0.85\linewidth]{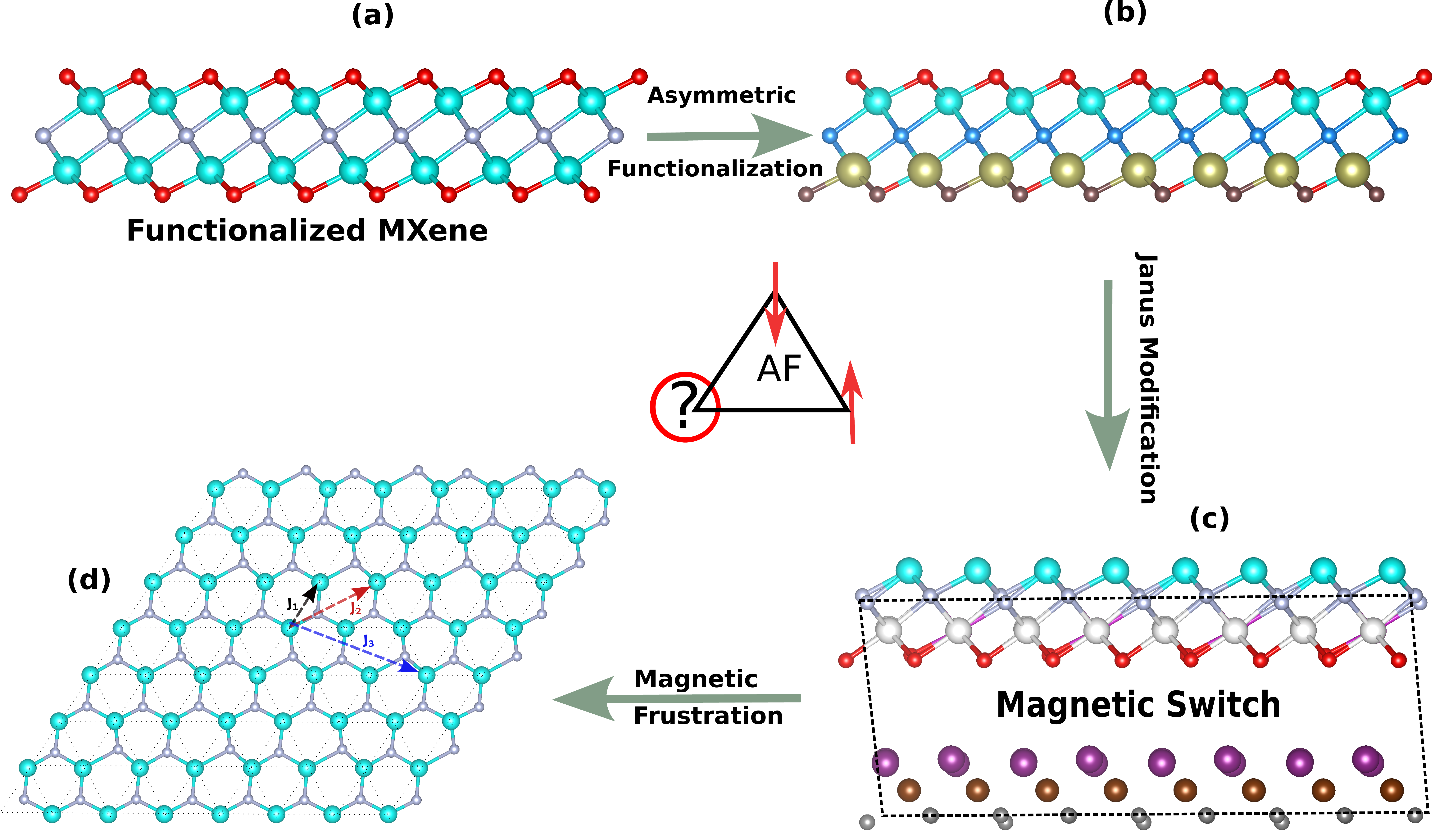}
\caption{Schematic representation of constructing geometrically frustrated MXene. 
(a) Functionalized MXene on both sides. 
(b) Asymmetric functionalization of MXene. 
(c) Janus modification to attain a target layer that remains magnetic, while the rest are magnetically switched off by suitable functionalization. 
(d) Magnetically frustrated layer.}
\label{device}
\end{figure*}

Fig.~\ref{structure} shows the optimized structure of Janus MXene MnHf$_2$C$_2$FI. It belongs to the M$_3$X$_2$ family ($n=2$ in M$_{n+1}$X$_n$T$_x$) and consists of a seven-layer slab with the stacking sequence F--Hf--C--Hf--C--Mn--I along $z$. The three metal layers are Hf, Hf and Mn, separated by two carbon layers, while the two outer surfaces are terminated by different halogens, F on the Hf side and I on the Mn side. All atoms lie on the three-fold axes of the hexagonal lattice and occupy the three sublattice positions in the order F(A)--Hf(C)--C(B)--Hf(A)--C(B)--Mn(C)--I(A).
The optimized in-plane lattice constant is $a=3.33$~\AA. Since the unit cell contains a single Mn atom, the Mn atoms form a two-dimensional triangular lattice with nearest-neighbor distance $a=3.33$~\AA\ and next-nearest-neighbor distance $\sqrt{3}\,a=5.76$~\AA, corresponding to the distances of the $J_1$ and $J_2$ interactions in the spin model. The Mn atom bonds to three C atoms below it at 2.24~\AA\ and to three I atoms above it at a much longer distance of 3.07~\AA, reflecting the large size of iodine. On the opposite surface, F bonds to three Hf atoms at 2.36~\AA. The Hf--C bond lengths are 2.21, 2.53, and 2.31~\AA\ from the bottom to the top of the slab. The distance between the outermost F and I atoms is 8.93~\AA, and the vacuum region is approximately 20~\AA\ thick, which is sufficient to decouple periodic images.
The different atomic sizes and electronegativities of F and I make the two surfaces to relax differently and break the out-of-plane mirror symmetry. The structure has the polar space group $P3m1$ (No.~156, point group $C_{3v}$), as determined with \textsc{findsym}~\cite{stokes2005findsym}, with all atoms occupying the three-fold axes. Inversion symmetry is therefore broken, giving rise to an intrinsic out-of-plane electric dipole and, together with spin--orbit coupling, allowing a finite Dzyaloshinskii--Moriya interaction.
The dynamical stability of the Janus MnHf$_2$C$_2$FI monolayer was assessed from its $\Gamma$-point vibrational spectrum [Table II, Supplementary Material]. The absence of imaginary frequencies supports its dynamical stability.

\subsection{Topological Study of Multi-Meronic States in Janus MXenes}
A minimal model of geometric frustration is discussed in Supplementary Material, Section II. To bring those theoretical ideas into a physical material system and move beyond earlier model-based approaches as discussed in Supplementary Material, we propose a Janus construction designed to engineer geometric frustration and host nontrivial magnetic topologies. Functionalizing two-dimensional (2D) MXenes provides a versatile route for tuning their structural, electronic, and magnetic properties\cite{akgenc2021tuning,chen2023two,amrillah2023potential,garcia2023theoretical,ram2024tuning,helmi2025novel,qian2025biomimetic,thasitha2025janus,limbu2025magnetic,lee2026asymmetrically,faysal2026first,kim2026magnetic}. In particular, asymmetric surface termination produces a Janus configuration with chemically distinct top and bottom surfaces, as schematically illustrated in Fig.~\ref{device}(a)-(d). The resulting structural inversion asymmetry, together with strong spin--orbit coupling (SOC), generates a macroscopic asymmetric crystal field. This can enable antisymmetric exchange interactions, most notably the Dzyaloshinskii--Moriya interaction (DMI), provided the relevant crystal site symmetries are satisfied. In addition, orbital hybridization between the transition-metal $d$ states and the valence orbitals of the heavier surface-terminating atoms can further enhance these relativistic effects, thereby providing an effective way to tune the underlying magnetic exchange interactions\cite{aldarawsheh2022emergence,abuawwad2026chiral,abuawwad2023crte}.

\begin{figure}
\includegraphics[width=0.94\linewidth]{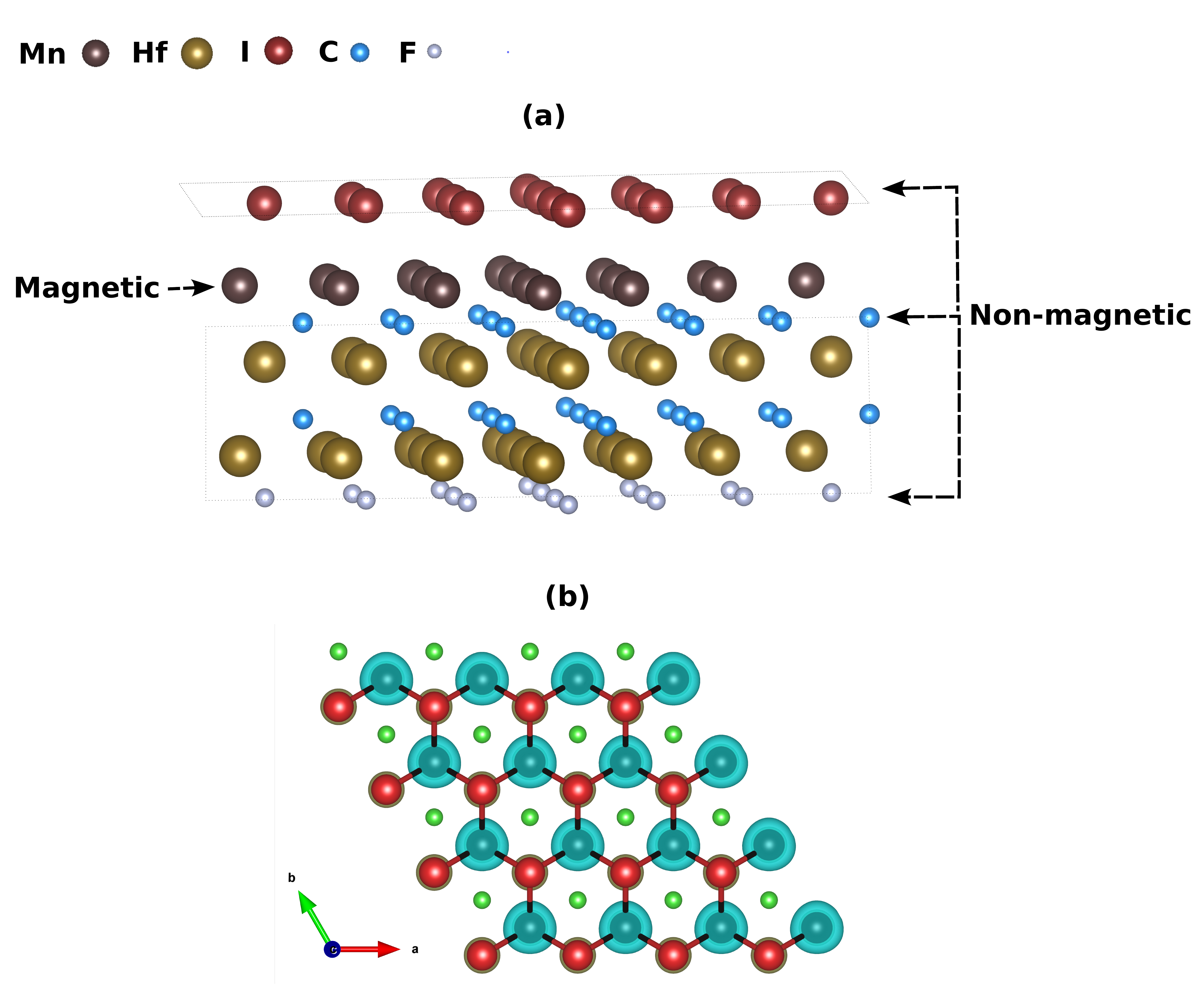}
\caption{(a) Illustration of Janus MnHf$_2$C$_2$FI with magnetic and non-magnetic layer. (b) Spin density showing Mn center only has significant spin density.}
\label{mag_non_spin_density}
\end{figure}

\begin{figure*}
\includegraphics[width=0.85\linewidth]{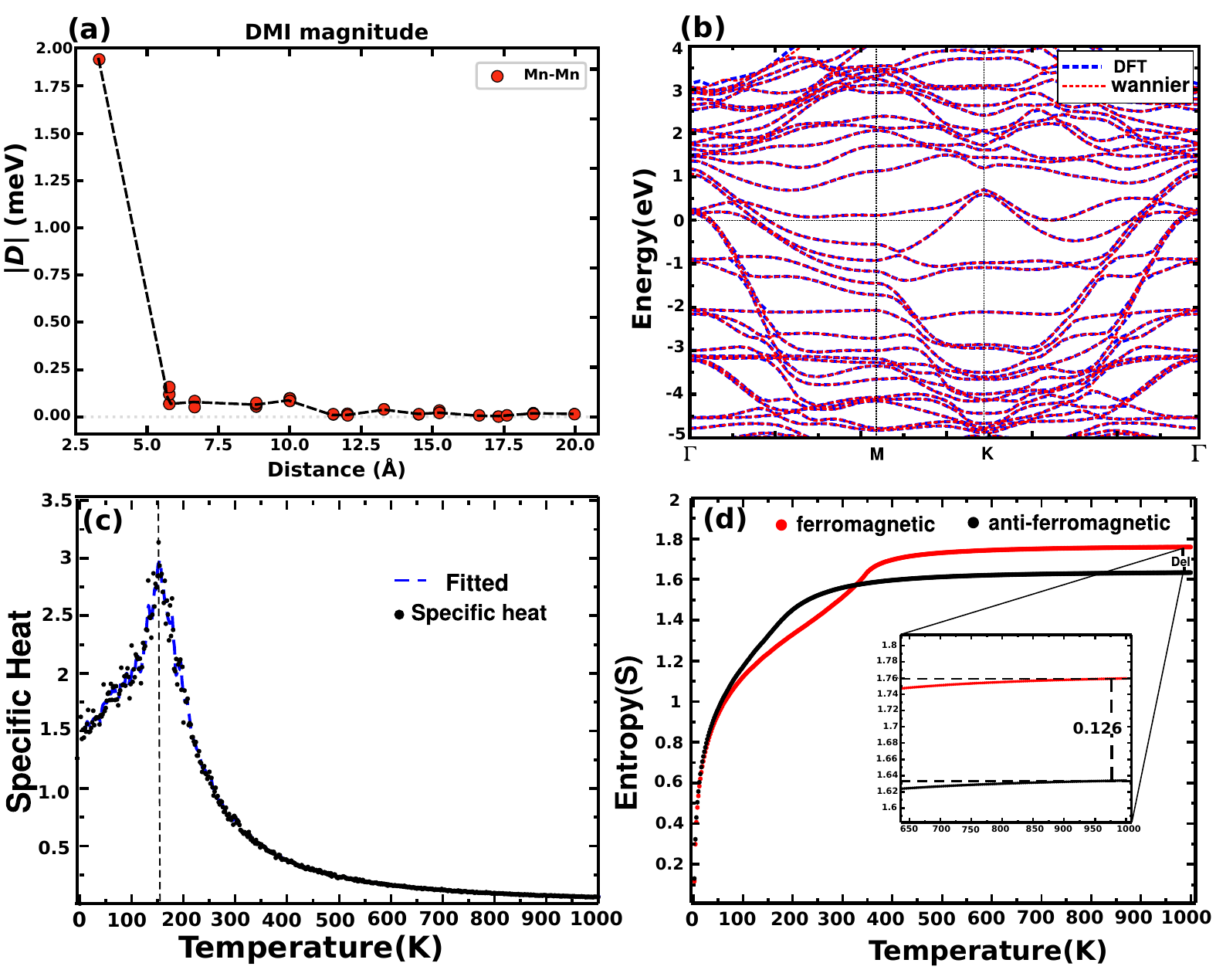}
\caption{The magnitude of the DMI vector, \( |\mathbf{D}_{ij}| \), is shown as a function of the Mn–Mn interatomic distance. Different markers at the same distance indicate distinct Mn–Mn bond interactions associated with different crystallographic directions. (b) Comparison of the DFT and Wannier-interpolated band structures, for constructing Wannier Hamiltonian used to calculate the magnetic exchange interactions.
(c) Magnetic specific heat as a function of temperature.
(d) Entropy obtained by integrating $C_v/T$ for the frustrated Hamiltonian and for a ferromagnetic reference; the high-temperature difference is the entropy deficit discussed in Sec.~\ref{sec:entropy}.}
\label{MnHfCFI_residual}
\end{figure*}

\begin{figure*}
\includegraphics[width=0.9\linewidth]{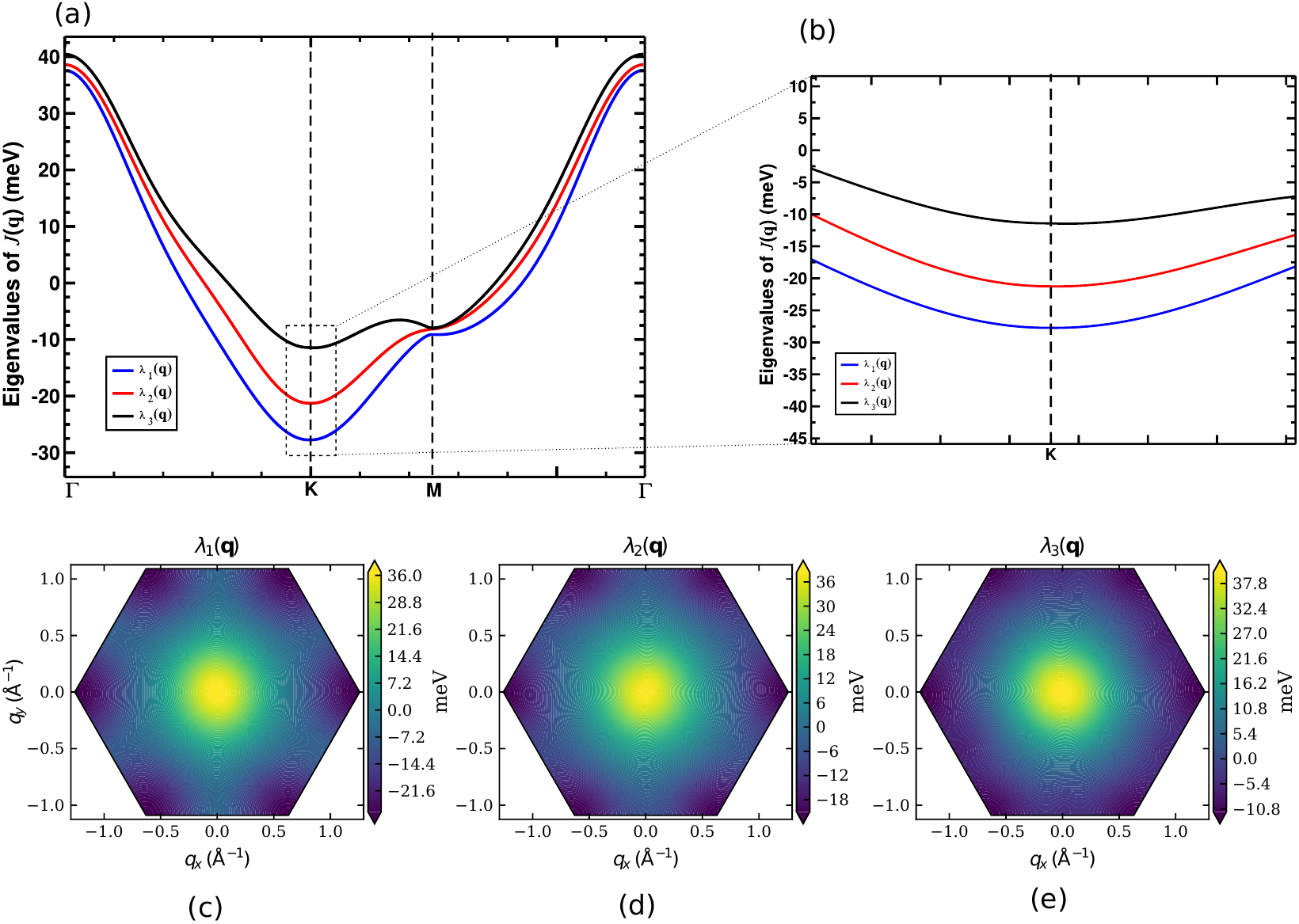}
\caption{Momentum-dependent eigenvalue spectrum of the reciprocal-space exchange matrix. (a) Eigenvalue $J(\mathbf q)$ dispersion along the high-symmetry path $\Gamma\!-\!K\!-\!M\!-\!\Gamma$. (b) Zoomed view of region (a) marked with a dotted box. (c)--(e) Corresponding two-dimensional contour maps of the eigenvalues $\lambda_1(\mathbf q)$, $\lambda_2(\mathbf q)$, and $\lambda_3(\mathbf q)$, respectively, across the reciprocal-space region. Together, the dispersion and contour maps highlight the momentum-dependent exchange interactions and the locations of the eigenvalue extrema.}
\label{Jq}
\end{figure*}

These interactions, competing with ferromagnetic (FM) and antiferromagnetic (AFM) exchange couplings on a geometrically frustrated lattice, can give rise to complex magnetic ordering and nontrivial spin textures. Our calculations reveal a pronounced layer-selective magnetic response in the Janus MXene, magnetism is localized predominantly in the Mn layer, whereas the remaining layers are essentially nonmagnetic (see Fig.\ref{mag_non_spin_density}(a)-(b)). This spatially selective magnetism highlights the potential of Janus functionalization to regulate the distribution of magnetic moments within a layered material. The observed magnetic–nonmagnetic contrast suggests a possible route to layer-selective magnetic control, similar to a single monolayer of magnetic elements expected to be antiferromagnetic on initially nonmagnetic surfaces.\cite{gao2008revealing,aldarawsheh2023intrinsic,rodriguez2024phase,sozeri2026frustrated}. To better understand the interplay among geometric frustration, symmetric exchange interactions, spin--orbit coupling (SOC)-mediated Dzyaloshinskii--Moriya interactions (DMI), and magnetocrystalline anisotropy in stabilizing unconventional magnetic configurations, we systematically investigated how competing interactions within the Mn layer work together to favor noncollinear states. The results are depicted in Fig.~\ref{MnHfCFI_residual}.

Fig.~\ref{MnHfCFI_residual}(a) shows the magnitude of the DMI vector, \( |\mathbf{D}_{ij}| \), as a function of the Mn--Mn interatomic distance, with individual bond contributions presented in Fig.~S11. The DMI is strongest for the nearest-neighbour Mn--Mn interaction at approximately 3.33~\AA{}, with a magnitude of about 1.95~meV, corresponding to approximately 36\% of the nearest-neighbour isotropic exchange interaction. The DMI decreases substantially with increasing Mn--Mn separation, with all longer-range contributions remaining below 0.22~meV. Although the DMI-to-exchange ratio becomes relatively large for the second-neighbour interactions at approximately 5.8~\AA{}, this primarily originates from the near-vanishing second-neighbour isotropic exchange rather than an enhancement of the DMI. Together with the dominant antiferromagnetic nearest-neighbour exchange (\(J_1 \approx -5.52\)~meV and the weaker antiferromagnetic third-neighbour coupling (\(J_3 \approx -0.7\) to \(-0.8\)~meV), these competing interactions provide the basis for frustrated noncollinear magnetic ordering.
 
Fig.~\ref{MnHfCFI_residual}(b) displays the Wannier-interpolated electronic band structure used to extract these microscopic magnetic exchange parameters (see all six direction Wannier interpolated plot in Fig.~S12). The Wannier model reproduces the DFT band structure well over the whole energy window. Note that the system is metallic, with several bands crossing the Fermi level. The specific heat in Fig.~\ref{MnHfCFI_residual}(c) shows a pronounced peak at $T^{*}\approx150$~K, which marks the build-up of $120^\circ$ N\'eel correlations in the Mn layer. Since the layer is two-dimensional and the anisotropy is weak, we refer to $T^{*}$ as an ordering crossover rather than a sharp N\'eel transition. Fig.~\ref{MnHfCFI_residual}(d) compares the entropy obtained by integrating $C_v/T$ for the actual Hamiltonian with that of a ferromagnetic reference. The frustrated magnet releases about $0.126\,k_{\mathrm B}$ per spin less entropy on heating. For classical vector spins this deficit is not a ground-state degeneracy, as it is for the Ising models of the Supplementary Material; it reflects instead that the frustrated magnet already carries more entropy at low temperature because its spin-wave spectrum is softer, in line with the flat $J(\mathbf q)$ landscape around K discussed below. It is therefore a thermodynamic measure of frustration (Sec.~\ref{sec:entropy}).

\subsection{Entropy deficit as a measure of frustration}\label{sec:entropy}
To quantify how frustrated the Mn layer is, as shown in the thermodynamics of Fig.~\ref{MnHfCFI_residual}(d). Relative to the ferromagnetic reference, the frustrated Hamiltonian releases $0.126\,k_{\mathrm{B}}$ per spin less entropy on heating. For the Ising models of the Supplementary Material, the same construction gives the residual entropy, e.g. $0.323\,k_{\mathrm B}$ per spin for the nearest-neighbor triangular antiferromagnet, and its reduction with $J_2$ tracks the lifting of the ground-state degeneracy. For classical vector spins the meaning is different. The $120^\circ$ ground state is unique up to global rotations, so there is no extensive ground-state degeneracy; the deficit measures instead how much softer the low-energy spectrum of the frustrated magnet is, since soft modes carry entropy already at low temperature. In this sense it plays the role that the residual entropy plays in the Ising case: it is a thermodynamic measure of frustration, and its moderate value is consistent with the flat but non-degenerate $J(\mathbf q)$ landscape around K.
The magnetic moment of $4.33\,\mu_{\mathrm B}$ per Mn obtained from DFT, together with the projected density of states (Fig.~S15), points to a high-spin, Mn$^{2+}$-like $3d^5$ configuration, reduced from $5\,\mu_{\mathrm B}$ by hybridization with the C and I states.

To characterize the spin spirals, we performed a Fourier transformation of the extracted magnetic interaction tensor J$_{ij}$ using the equation (see Appendix~\ref{FourierJq}). In Fig.~\ref{Jq}(a), the global minimum lies at the K point, so the ground state is the $120^\circ$ N\'eel state of the triangular lattice. Around K, however, the landscape is rather flat (Fig.~\ref{Jq}(b)): modulations close to the N\'eel state cost little exchange energy, and this is what allows the textures discussed below to exist as metastable states. Note that Fig.~\ref{Jq}(a) shows three dispersion curves corresponding to spin spirals propagating along high-symmetry directions in the Brillouin zone. This apparent band splitting is due to the lifting of the degeneracy of the Fourier-transformed exchange tensor J(\textbf{q}) by the antisymmetric DMI interaction and the anisotropic exchange tensor. The entire first Brillouin zone was also scanned to pinpoint minima away from the high-symmetry axes in Fig.~\ref{Jq}(c-e), corresponding to the three bands, or eigenvalues, $\lambda_1(\mathbf q)$, $\lambda_2(\mathbf q)$, and $\lambda_3(\mathbf q)$, respectively. It also shows that the minima of the three bands are at the K point. Among the three, $\lambda_1(\mathbf q)$ is the lowest-energy band; its minimum at K selects the $120^\circ$ N\'eel ground state of MnHf$_2$C$_2$FI, with a vector chirality fixed by the DMI.

\begin{figure*}
    \centering
    \includegraphics[width=\linewidth]{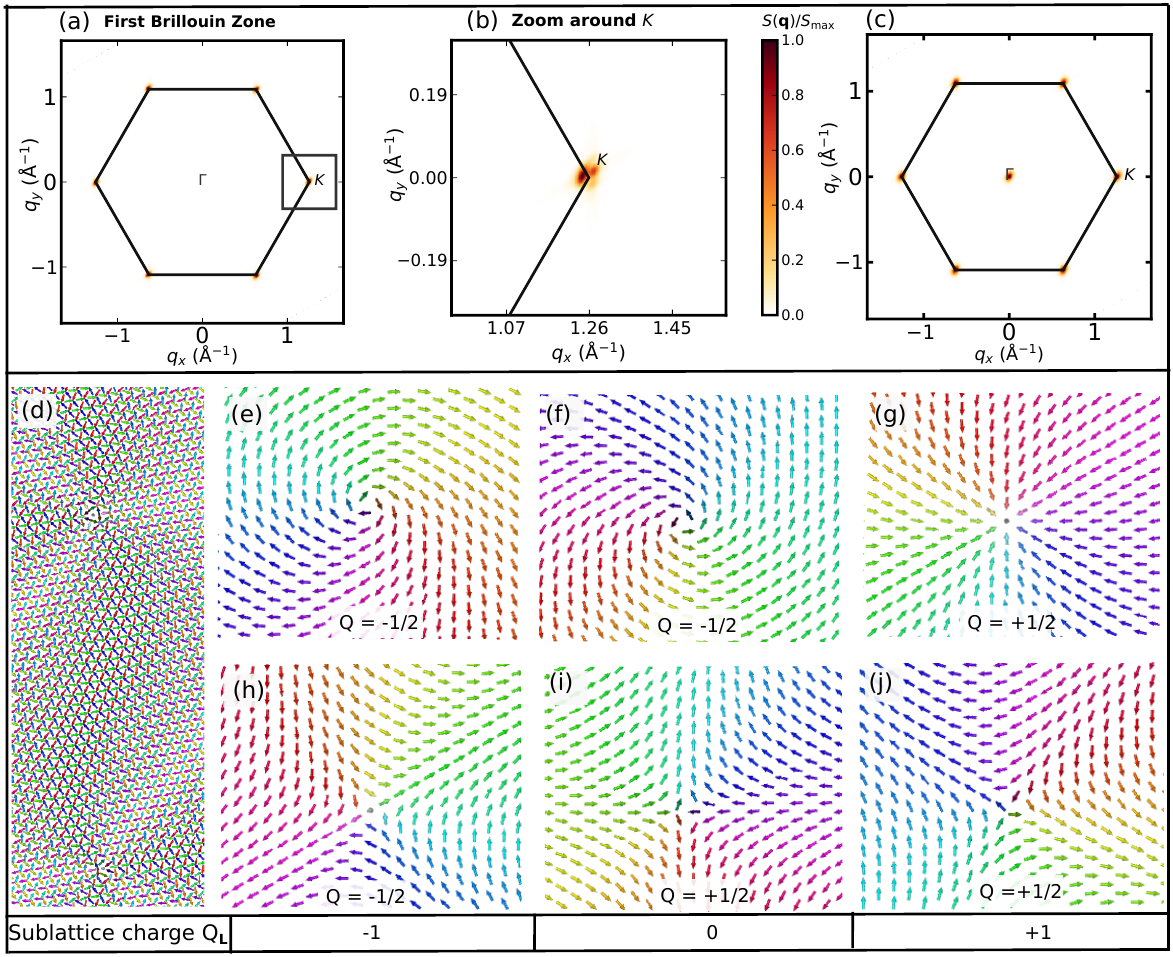}
\caption{Normalized static spin structure factor \(S(\mathbf{q})/S_{\mathrm{max}}\) across the first hexagonal Brillouin zone (a) and around the \(K\) point (b). The spectral weight is concentrated at the six \(K\) corners, with no central \(\Gamma\)-point peak, indicating zero net magnetization. The zoomed view reveals an anisotropic, broadened intensity profile around \(K\), reflecting the long-wavelength spatial modulation of the hexameronic cluster. (c) All three magnetic sublattices exhibit identical static structure factors, confirming the common \(\sqrt{3} \times \sqrt{3}\) (\(120^{\circ}\)) periodicity and long-wavelength topological modulations. Sharp peaks at the six \(K\) corners reflect dominant three-sublattice \(120^{\circ}\) antiferromagnetic order, while the central \(\Gamma\) peak reflects smooth long-wavelength spin modulation and a net uniform magnetization component within each sublattice. (d) Antiferromagnetic (AFM) hexameronic state with a localized vortex–antivortex texture in the geometrically frustrated triangular Mn layer. (e)–(j) Decomposition of the AFM texture into six FM-like meronic constituents forming three vortex–antivortex pairs associated with the three magnetic sublattices. Q$_L$ is the total charge of each sublattice.}
\label{multimeron}
\end{figure*}

\begin{figure*}
    \centering
    \small
    \includegraphics[width=0.97\linewidth]{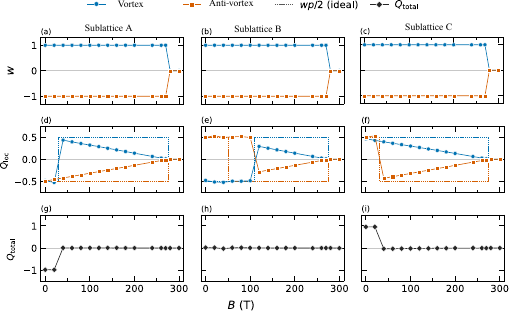}
\caption{Field evolution of the sublattice-resolved topology of the hexameron. The columns correspond to sublattices A, B, and C. Panels (a)–(c) show the winding numbers $w$ of the vortex (blue circles) and antivortex (orange squares) on each sublattice. Panels (d)–(f) show the local topological charges $Q_{\mathrm{loc}}$, obtained by summing discrete spin solid-angle contributions within a radius of $2.5$~nm around each selected core. Dashed curves indicate the ideal meron values $wp/2$, where $p$ is the core polarity. Panels (g)–(i) show the integrated topological charge $Q_{\mathrm{total}}$ of each sublattice. Core-charge reversals occur without changes in winding, revealing internal topological rearrangements within the hexameron. All six tracked winding numbers reach zero between approximately $270$ and $280$~T, accompanied by vanishing local charges.}
\label{topology_vs_mag}
\end{figure*}

To gain microscopic insight into these phases, we performed atomistic spin simulations and computed the static spin structure factor, which allows us to examine both the real-space ground-state spin configurations and their reciprocal-space ($\mathbf{k}$-space) signatures. Besides the uniform $120^\circ$ N\'eel ground state, the spin relaxations reveal localized hexameronic textures embedded in the $120^\circ$ background, as shown in Fig.~\ref{multimeron}(d) [also see Fig.~S13]. The hexameron is not the ground state: its energy lies above that of the uniform N\'eel state by $\Delta E=$~250.82~meV per hexameron. It is, however, a local minimum of the energy landscape, as confirmed by the strictly positive spin-wave spectrum in Sec.~\ref{sec:magnons}, and therefore a metastable state. The texture consists of a vortex--antivortex pair centered on Mn lattice sites, with surrounding spins forming meronic configurations. These textures arise from magnetic frustration associated with the underlying antiferromagnetic interactions. The resulting hexameronic state, characterized by a vortex--antivortex pair within the geometrically frustrated triangular Mn sublattice, can be further decomposed into six constituent FM-like vortex--antivortex units, as illustrated in Fig.~\ref{multimeron}(e)--(j). These configurations are characterized by the winding number $w$ and the local meron number $Q_{\mathrm{loc}}=wp/2$. The winding number is obtained by integrating counterclockwise along a contour $C$ that encloses a single core\cite{koshibae2016berry}:
\begin{equation}
w=\frac{1}{2\pi}\oint_C d\phi,
\end{equation}
where $\phi=\operatorname{atan2}(S^y,S^x)$ denotes the azimuthal angle of the in-plane spin component. With this convention, a vortex has $w=+1$, whereas an antivortex has $w=-1$. The core polarity is defined as $p=+1$ ($p=-1$) for an up (down) out-of-plane core magnetization. Thus, each meronic component carries a local topological charge $Q_{\mathrm{loc}}=wp/2=\pm 1/2$.
Although the individual meron and antimeron components carry half-integer local topological charges, their combined arrangement gives integer topological charges when the discrete charge is integrated separately over each magnetic sublattice. The three sublattices yield $Q\simeq-1$, $Q\simeq0$, and $Q\simeq+1$, respectively [see Fig.~S14(d)--(f)], corresponding to one ferromagnetic-like meronic pair on each sublattice. The sublattice charges satisfy $Q_1+Q_2+Q_3\simeq0$, indicating a compensation of the sublattice-resolved topological charges in the antiferromagnetic texture\cite{rybakov2019chiral}.

The calculated spin structure factor of the corresponding spin states, as shown in Fig.~\ref{multimeron}(a). Unlike canted or ferrimagnetic states, which show finite spectral weight at the zone center, the complete absence of a $\Gamma$-point peak confirms full spin compensation across the entire supercell: the background $120^{\circ}$ sublattice moments cancel in-plane, while the oppositely polarized $S_z = \pm 1$ meron and antimeron cores cancel out-of-plane. As a result, the magnetic spectral weight is concentrated around the six $K$ points. More importantly, zooming in on the reciprocal-space distribution around $K$ in Fig.~\ref{multimeron}(b) shows that the sharp Bragg delta function of a uniform $120^{\circ}$ state is replaced by an anisotropic, broadened satellite structure. This splitting occurs because the bound hexameron cluster imposes a long-wavelength spatial phase modulation $e^{i\theta(\mathbf{r})}$ on the underlying $\sqrt{3} \times \sqrt{3}$ order, shifting spectral weight to satellite wavevectors $\mathbf{q}_K \pm \Delta\mathbf{q}$ set by the superlattice period $L_{\mathrm{hex}}$ ($\Delta q \sim 2\pi/L_{\mathrm{hex}}$), while the composite six-defect geometry breaks local rotational symmetry around $K$.
To isolate the topological structure of the hexameron from the rapid $\sqrt{3} \times \sqrt{3}$ background modulation, we project the spin configuration onto a single sublattice and compute its corresponding static spin structure factor $S(\mathbf{q})/S_{\mathrm{max}}$ [Fig.~\ref{multimeron}(c)].
In contrast to the full three-sublattice structure factor where uniform magnetization cancels out, the single-sublattice $S(\mathbf{q})/S_{\mathrm{max}}$ in Fig.~\ref{multimeron}(c) exhibits a prominent peak at the zone center $\Gamma$ alongside peaks at the six $K$ corners. The emergence of strong $\Gamma$-point spectral weight reflects the locally smooth, nearly parallel alignment of spins within an individual sublattice across the regions separating the defect cores, whereas the $K$-point peaks capture the periodic lattice modulation and the spatial arrangement of the six meron and antimeron cores.

To separate the out-of-plane core polarizations from the coplanar $120^{\circ}$ antiferromagnetic background, we resolve the static spin structure factor, $S(\mathbf{q}) = \frac{1}{N}\sum_{\alpha=x,y,z}|S^{\alpha}(\mathbf{q})|^2$, into transverse in-plane [$S^{\perp}(\mathbf{q}) = S^{xx}(\mathbf{q}) + S^{yy}(\mathbf{q})$] and longitudinal out-of-plane [$S^{zz}(\mathbf{q})$] components (see Fig.~S17). In the composite hexameron state, the spectral weight is overwhelmingly concentrated in the in-plane spin components, with an intensity ratio of $S^{zz}_{\mathrm{max}}/S^{\perp}_{\mathrm{max}} \approx 6.1 \times 10^{-4}$, showing that the out-of-plane canting is strongly confined to the defect cores. Notably, $S^{zz}(\mathbf{q})$ displays its dominant Bragg peaks at the six $K$ high-symmetry points rather than at the zone center $\Gamma$, indicating that the out-of-plane core magnetizations follow the $\sqrt{3} \times \sqrt{3}$ periodicity across the three magnetic sublattices. To identify the origin of this periodicity, we project the spin texture onto its three constituent $\sqrt{3} \times \sqrt{3}$ sublattices, each of which hosts a pair of fractional topological defects, for a total of six meron and antimeron cores within the supercell [Figs.~\ref{multimeron}(e)--(j)]. Since the three sublattices are related by global $120^{\circ}$ spin-space rotations and primitive lattice translations, their individual structure factors are nearly the same. Within each sublattice, the collinear ferromagnetic background between the two defect cores produces a strong in-plane Bragg peak at $\Gamma$ together with the six $K$-point reflections, as well as a weak residual longitudinal intensity, $S^{zz}(\Gamma) \sim 3.7 \times 10^{-3}\text{--}1.7 \times 10^{-1}$, arising from minor core-profile asymmetries between the $Q_{\mathrm{local}} = \pm \frac{1}{2}$ defects. When the three sublattices are recombined, destructive interference from the $120^{\circ}$ phase relation ($1 + e^{i 2\pi/3} + e^{i 4\pi/3} = 0$) eliminates the in-plane $\Gamma$-point weight and suppresses $S^{zz}(\Gamma)$ to $1.3 \times 10^{-5}$, confirming complete magnetic compensation across the hexameron supercell.

\begin{figure*}
\centering
\includegraphics[width=0.97\linewidth]{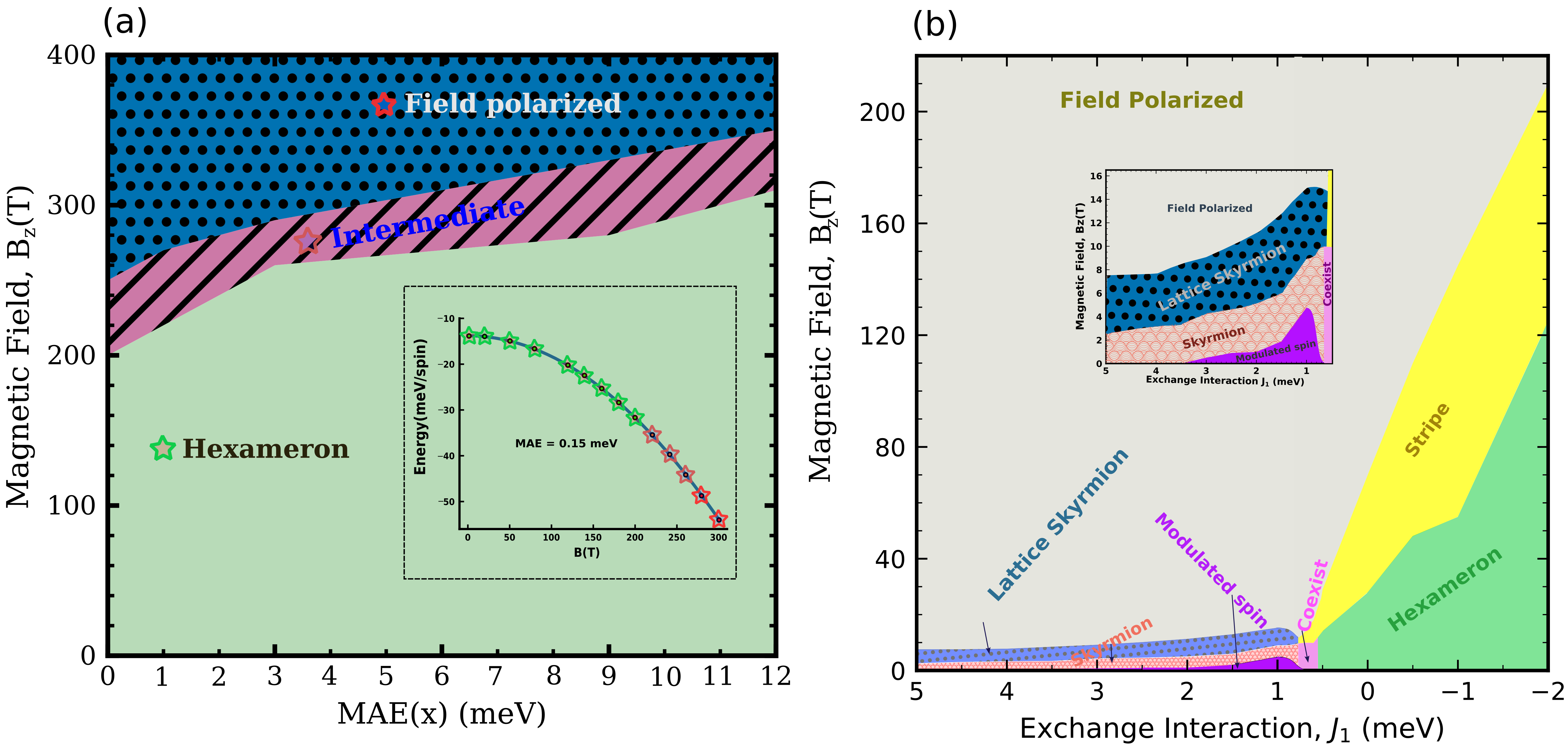}
\caption{
(a) Stability diagram of the metastable hexameron as a function of the in-plane anisotropy $K$ (easy axis along $x$) and the out-of-plane field $B$, showing where the hexameron survives, the intermediate regime, and the field-polarized state. Inset: Energy per spin $E$ as a function of external magnetic field $B$ from \qty{0.0}{\tesla} to \qty{300.0}{\tesla} for $MAE = 0.15$meV. Markers show the data points, and the solid curve indicates the continuous parabolic trend.
The quadratic decrease observed above $B > \qty{20}{\tesla}$ suggests progressive Zeeman alignment and field-induced spin polarization. (b)  Calculated magnetic phase diagram of the Janus MXene monolayer as a function of only the first-neighbor exchange interaction $J_1$ and the out-of-plane magnetic field $B$. The phase space is divided into a high-$J_1$ chiral skyrmion regime ($J_1 \ge 1.0$\,meV) and a low/frustrated negative-$J_1$ regime dominated by hexamerons and stripe domains ($J_1 \le 0.5$\,meV). A smooth sigmoidal crossover boundary outlines the topological coexistence window ($0.55\,\text{meV} \le J_1 < 0.77\,\text{meV}$) where modulated/stripe domains coexist alongside a frustrated Néel state[see Fig.~S16 for all real-space spin vector configurations(phases)]. The inset shows an enlarged view of the low-field region.
}
\label{phase_diagram}
\end{figure*}

We also investigated the field-dependent topology of this hexameron. It is observed that [Fig.~\ref{topology_vs_mag}(a)--(i)] there is an internal reorganization of the six-core structure, with one vortex--antivortex pair on each of the three magnetic sublattices. The six constituent cores retain their winding numbers, $w=+1$ or $-1$, up to approximately $270$~T, even as their local topological charges change[Fig\ref{topology_vs_mag} (a)-(c)]. At low fields, the integrated sublattice local charges are $(Q_A,Q_B,Q_C)\approx(-1,0,+1)$, reflecting compensation between topologically distinct sublattice textures. Interestingly, the positive and negative local topological charges within the hexameron strongly compensate throughout the applied field, yielding a zero total charge of three sublattices, $Q_{\mathrm{tot}}=\sum_i Q_{\mathrm{loc},i}\simeq 0$, despite pronounced variations in the individual contributions[Fig\ref{topology_vs_mag} (d)-(f)]. This compensation suggests that the local topological structure can reorganize while preserving the global topological neutrality of the hexameron[Fig\ref{topology_vs_mag} (g)-(i)].

On sublattice A, the vortex-associated charge changes sign between approximately $20$ and $40$~T, and the antivortex-associated charge on C does the same, leaving each sublattice with nearly vanishing integrated charge. The two core-associated charges on B then reverse between approximately $100$ and $120$~T, while $Q_B$ remains close to zero. These changes are consistent with sequential core-polarity reversals, as indicated by the $wp/2$ reference curves, rather than vortex--antivortex conversion. The hexameron therefore preserves its six-core winding structure while accommodating different internal charge configurations.

For an out-of-plane field, the Zeeman interaction favors spin components aligned with the field and competes with the interactions that stabilize the noncollinear hexameron. The different reversal fields reflect the sublattice-dependent nature of this competition. After the reversals, the magnitudes of the local charges, evaluated within $2.5$~nm of each core, decrease while the winding numbers remain unchanged. This distinction arises because winding measures in-plane spin rotation, whereas topological charge also depends on the out-of-plane spin profile. For an axisymmetric constituent,
$Q(R)=w[\cos\theta(0)-\cos\theta(R)]/2$;
field-induced changes in the boundary canting can therefore reduce $Q(R)$ without changing $w$. Accordingly, a departure from the ideal meron value $wp/2$ does not by itself imply loss of the constituent cores. Between approximately $270$ and $280$~T, all six tracked winding numbers become zero, along with the local charges, indicating the disappearance of the resolved hexameron winding structure. Thus, the simultaneous loss of winding and local charge in this field range marks the field-induced disappearance of the resolved hexameron winding structure.

\subsection{Field and anisotropy stability of the hexameron}
Fig.~\ref{phase_diagram}(a) shows where the metastable hexameron survives when we vary the magnetic anisotropy energy (MAE) along the easy axis $x$ and the external magnetic field $B$ along the $ z$ direction. Starting from the hexameron and increasing the field, we find three regimes: the hexameron itself, an intermediate field-induced state, and the field-polarized state. Since the hexameron is metastable, the boundaries below are stability limits of the texture, not thermodynamic phase boundaries. For a given value of the MAE, these regimes can be expressed as
\begin{equation}
\label{phase_regions}
\begin{aligned}
B &\leq B_{\mathrm{hex}}(K)
&& \text{(hexameron phase)},\\
B_{\mathrm{hex}}(K) &< B < B_{\mathrm{pol}}(K)
&& \text{(intermediate phase)},\\
B &> B_{\mathrm{pol}}(K)
&& \text{(field-polarized phase)}.
\end{aligned}
\end{equation}
where $K$ denotes the MAE, $B_{\mathrm{hex}}$ is the upper stability field of the hexameron phase, and $B_{\mathrm{pol}}$ is the onset field of the fully polarized state.
At zero MAE, the hexameron configuration persists up to $B_{\mathrm{hex}}=200$~T. Increasing the field beyond this value produces an intermediate phase in the interval $200<B<250$~T, followed by complete field polarization for $B>250$~T. This field scale is set by the antiferromagnetic exchange rather than by a special robustness of the texture: for a classical triangular antiferromagnet the saturation field is $B_{\mathrm{sat}}\simeq9|J_1|/\mu_{\mathrm{Mn}}$, which gives about 200~T for $J_1\approx-5.52$~meV and $\mu_{\mathrm{Mn}}=4.33\,\mu_{\mathrm B}$, close to the values found here. The stability range of the hexameron phase expands systematically with increasing MAE. In particular, $B_{\mathrm{hex}}$ increases from $200$~T at $K=0$~meV to $310$~T at $K=12$~meV. For the pristine lattice, our DFT calculations give a magnetocrystalline anisotropy energy (MAE) of 1.56 meV. To capture the effects of external perturbations such as uniaxial in-plane strain or substrate-induced symmetry breaking—which lifts the three-fold \(C_{3v}\) rotational symmetry and can significantly enhance anisotropy—we introduce an effective uniaxial anisotropy along the x-axis and explore K values up to 12 meV. This clearly shows that in-plane magnetic anisotropy opposes field-induced polarization and helps stabilize the noncollinear hexameron texture over a wider field range.
The polarization threshold exhibits a corresponding increase, rising from $B_{\mathrm{pol}}=250$~T at $K=0$~meV to $350$~T at $K=12$~meV. Thus, larger anisotropy energies require stronger magnetic fields to align the magnetic moments completely along the field direction. The intermediate phase occupies a field window of approximately $30$--$50$~T throughout the investigated MAE range. Its width does not vary monotonically with the MAE, suggesting that its stability is controlled by a nontrivial competition among the Zeeman interaction, magnetic anisotropy, and the interactions responsible for stabilizing the hexameron texture.
The two phase limits both shift toward higher magnetic fields as the MAE increases. This behavior demonstrates that the phase progression is governed by the competition between anisotropy and Zeeman’s energies. At low fields, the interactions stabilizing the hexameron texture dominate. At intermediate fields, the original texture is transformed into a distinct frustration-like magnetic configuration before complete polarization occurs~[see Fig.~S16 (g)]. At sufficiently large fields, the Zeeman contribution dominates and produces a nearly uniform state aligned with the applied field whose energy decreases quadratically with $B$ [see inset Fig.~\ref{phase_diagram}(a)].

\begin{figure}
    \centering
    \includegraphics[width=\linewidth]{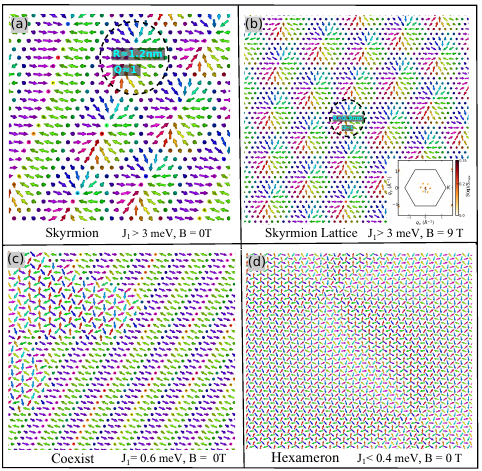}
    \caption{Evolution of the spin textures as a function of the nearest-neighbor exchange interaction $J_1$ and the applied magnetic field $B$. (a) Isolated skyrmion stabilized for $J_1>3$~meV at $B=0$~T, with a radius of $R=\SI{1.2}{nm}$ and a topological charge of $Q\approx 1$. (b) Skyrmion lattice stabilized for $J_1>3$~meV at $B=9$~T. (c) Coexistence of frustrated spin clusters and modulated stripe domains at $J_1=0.6$~meV and $B=0$~T. (d) Hexameronic state stabilized for $J_1<0.4$~meV at $B=0$~T. Colors represent the spin orientations.}
    \label{ferro_antiferro}
\end{figure}

To better understand how frustration controls the magnetic ground state, we constructed the phase diagram of the Janus monolayer as a function of the first-neighbor exchange $J_1$ and the out-of-plane magnetic field $B_z$ [Fig.~\ref{phase_diagram}(b)], while keeping all other interactions fixed at their first-principles values. The phase boundaries separate the $(J_1,B_z)$ plane into two regimes with markedly different field stability. For $J_1 \gtrsim 1.0$~meV, the ordered phases remain confined to a narrow low-field region ($B_z \lesssim 15$~T) [inset Fig~\ref{phase_diagram}(b)] beyond which the system becomes field polarized. Within this strip, increasing $B_z$ drives the sequence modulated spin $\rightarrow$ skyrmion $\rightarrow$ skyrmion lattice $\rightarrow$ field polarized. The corresponding real-space configurations show, respectively, periodic spin bands, isolated skyrmions with a full $2\pi$ winding of the in-plane angle, and a hexagonally ordered skyrmion array [Figs.~S16(a)--(c)]. The narrow extent of this region in field shows that these DMI-stabilized textures are fragile against the Zeeman energy. For $J_1 \lesssim 0.5$~meV, the phase diagram changes qualitatively: the low-field strip is replaced by a wedge-shaped region that opens as $J_1$ decreases and is occupied by the hexameronic and stripe phases. Real-space spin maps show that these states consist of short-period noncollinear order, namely vortex--antivortex pairs in the hexameron and alternating rows in the stripe state [Figs.~S16(e),(f)], with no long-wavelength modulation. Because they arise from competing interactions rather than DMI, their saturation field $B_{\rm sat}$ increases almost linearly with $|J_1|$, reaching about 120~T for the hexameron and exceeding 200~T for the stripe phase at $J_1=-2$~meV. The two regimes meet in a narrow coexistence window ($0.55 \le J_1 < 0.77$~meV), where frustrated clusters and modulated/stripe domains remain spatially separated within the same sample [Fig.~S16(d) or Fig.~\ref{ferro_antiferro} (c)]. This marks the crossover between DMI-driven and frustration-driven order. All phase boundaries converge to the field-polarized state at $B_{\rm sat}$ [Fig.~S16(h)], reached through a partially polarized intermediate state [Fig.~S16(g)]. At the DFT value $J_1\approx-5.52$~meV, the Janus monolayer lies deep in the hexameronic regime, consistent with the metastable hexamerons found at zero field. Overall, the hexameron exists over a broad region of parameter space rather than at a fine-tuned point.

Fig.~\ref{ferro_antiferro} illustrates how the magnetic phases evolve with the nearest-neighbor exchange $J_1$ and the applied field $B$. For large $J_1$ ($J_1>3$~meV), the competition between the Heisenberg exchange and the Dzyaloshinskii--Moriya interaction (DMI) stabilizes isolated skyrmions even at zero field [Fig.~\ref{ferro_antiferro}(a)]. Each skyrmion has a radius
$R\approx\SI{1.2}{nm}$ and a topological charge $Q\approx1$. When an out-of-plane field of
$B=\SI{9}{T}$ is applied, the isolated skyrmions are compressed into a densely packed
skyrmion lattice [Fig.~\ref{ferro_antiferro}(b)], where each unit cell
retains $Q\approx1$ with $R\approx\SI{0.9}{nm}$. The inset shows the calculated two-dimensional static spin structure factor $S(\mathbf{q})$, exhibiting sharp sixfold-symmetric Bragg peaks characteristic of long-range crystalline order.
Decreasing $J_1$ progressively destabilizes the skyrmion phase. At $J_1=0.6$~meV, the system enters a coexistence regime [Fig.~\ref{ferro_antiferro}(c)] as discussed earlier, where frustrated clusters coexist with modulated stripe domains, indicating that the two states are nearly degenerate and that this is the transition point. For $J_1<0.4$~meV, the hexameronic state spans the entire sample
[Fig.~\ref{ferro_antiferro}(d)]. The $J_1$-driven
crossover from a topologically nontrivial skyrmion phase to a frustrated
hexameronic phase therefore reflects the growing role of geometric frustration at $J_1 < 0$ and is consistent with the frustration-driven hexameric state.

This study also suggests that Janus MXene could also serve as a promising platform for topologically protected skyrmions in the ferromagnetic regime ($J_1 > 0$), if a phase transition from antiferromagnetic to ferromagnetic interactions can be induced through pressure-induced transitions or chemical substitution in d- and f-electron compounds, as achieve in some of the previous work\cite{wysokinski2019mechanism,dugu2026interplay,pan2024strain,friedemann2018quantum,taufour2016ferromagnetic,kaluarachchi2017tricritical,jeffries2016emergent,pospivsil2017switching,lengyel2015avoided}.

\subsection{Topological Magnon Excitations and Thermal Hall Conductivity}\label{sec:magnons}
\begin{figure*}
    \centering
    \includegraphics[width=\linewidth]{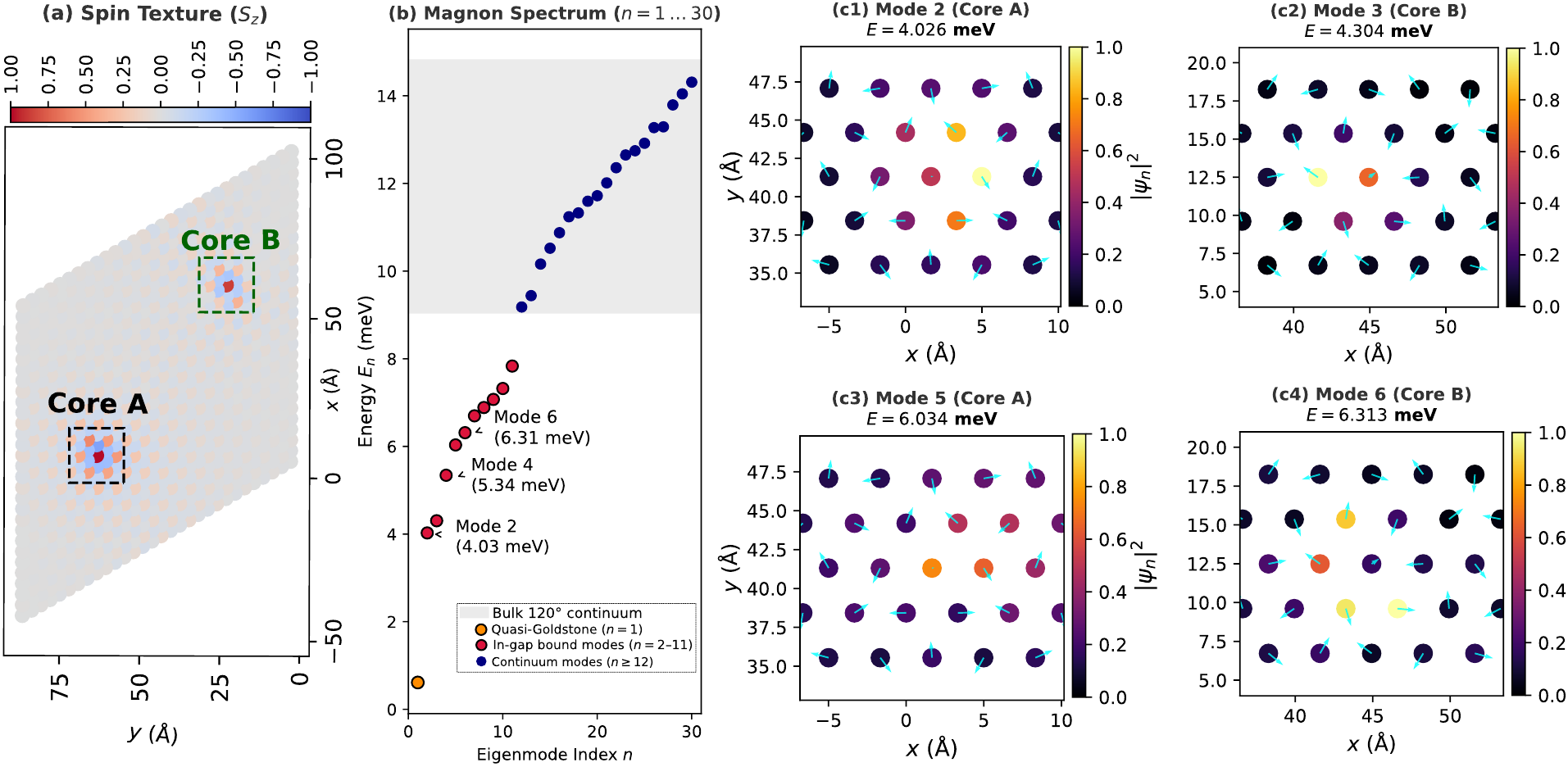}
    \caption{Real-space magnon localization in the frustrated hexameron state. (a) Out-of-plane spin component $S_z(\mathbf{r}_i)$ showing two hexameron cores (Core~A and Core~B, dashed boxes) embedded in the $120^\circ$ magnetic background. (b) Low-energy magnon eigenvalue spectrum $E_n$ ($n = 1 \dots 30$), highlighting the hybridized quasi-Goldstone mode ($n = 1$, $0.613~\mathrm{meV}$) and the discrete manifold of in-gap hexameron bound states ($n = 2 \dots 11$, $4.03$--$7.83~\mathrm{meV}$) separated from the bulk $120^\circ$ magnon continuum ($E \geq 9.18~\mathrm{meV}$, shaded region). (c1)--(c4) Zoomed-in spatial probability densities $|\psi_n(\mathbf{r}_i)|^2$ around Core~A and Core~B for representative bound modes ($n = 2, 3, 5, 6$), with cyan arrows indicating the local in-plane spin texture $(S_x, S_y)$.}
    \label{magnon_modes}
\end{figure*}

\begin{figure*}
    \centering
    \includegraphics[width=\linewidth]{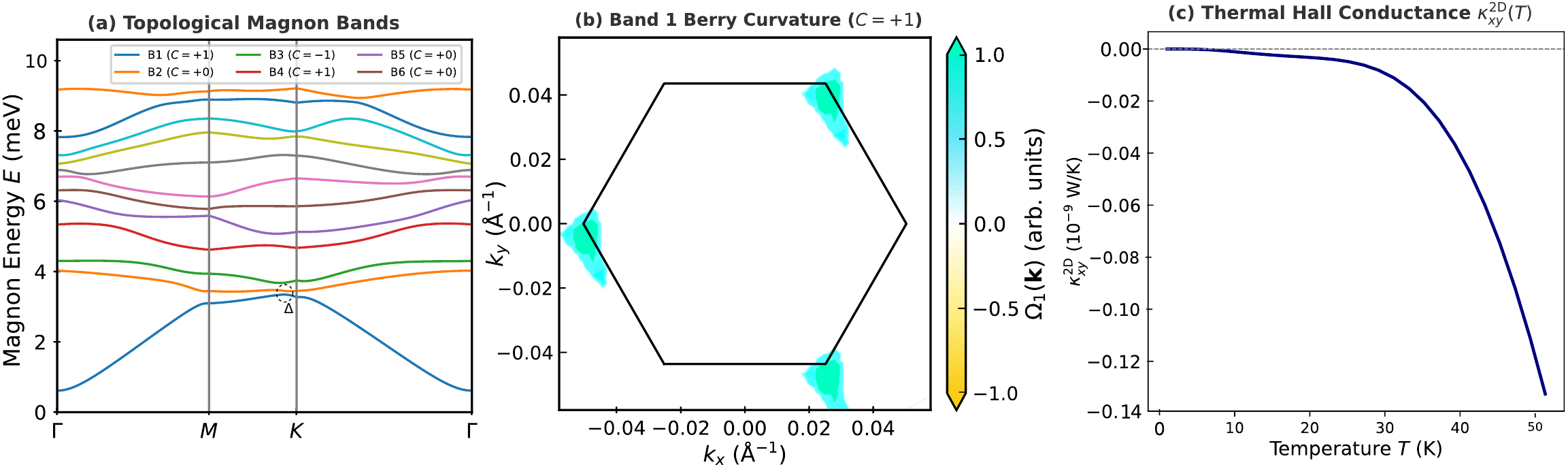}
    \caption{Momentum-space magnon topology and thermal Hall response. (a)~Magnon band dispersion along the high-symmetry path $\Gamma$--$M$--$K$--$\Gamma$, with the lowest branches labeled by their quantized Chern numbers $C_n$. (b)~Normalized Berry-curvature distribution $\Omega_1(\mathbf{k})$ for the lowest magnon branch ($C_1 = +1$) across the first hexagonal Brillouin zone (solid black boundary), showing $C_3$-symmetric hotspots at alternating $K$ vertices. (c)~Temperature dependence of the two-dimensional magnon thermal Hall conductance $\kappa_{xy}^{2\mathrm{D}}(T)$, showing a thermally activated negative response governed by the low-energy topological magnon bands.}
    \label{magnon_topology}
\end{figure*}
To test whether the hexameron is a true local minimum, and to see how its topology shows up in the excitations, we use linear spin-wave theory (LSWT) around the relaxed noncollinear texture. Starting from the first-principles magnetic exchange and Dzyaloshinskii--Moriya interactions mapped onto an effective Heisenberg--DMI spin Hamiltonian, we rotate the spin operators at each magnetic site $i$ into local quantization frames aligned with the classical ground-state orientation $
\mathbf{S}_i^{\mathrm{cl}}$ and carry out the Holstein--Primakoff expansion to quadratic order in $1/S$\cite{holstein1940field}:
\begin{equation}
    H_{\mathrm{LSWT}} = \frac{1}{2} \sum_{\mathbf{k}} \Psi_{\mathbf{k}}^\dagger \mathcal{H}_{\mathbf{k}} \Psi_{\mathbf{k}},
\end{equation}
where $\Psi_{\mathbf{k}} = \left(a_{\mathbf{k},1}, \dots, a_{\mathbf{k},M}, a_{-\mathbf{k},1}^\dagger, \dots, a_{-\mathbf{k},M}^\dagger\right)^T$ is the $2M$-component bosonic Nambu spinor spanning the $M$ magnetic sublattices in the supercell. The absence of linear bosonic terms (expansion residue $|\delta| < 10^{-13}$) shows that the hexameron is an exact classical stationary point, while para-unitary Bogoliubov--de~Gennes (BdG) diagonalization\cite{colpa1978diagonalization} gives strictly real, positive excitation energies $\omega_n(\mathbf{k}) > 0$, confirming that this stationary point is a local minimum. Together with $\Delta E>0$ relative to the N\'eel state, this establishes the hexameron as a metastable state. These \(k\)-space results were obtained using a \(30 \times 30\) supercell consisting of \(900\) lattice sites.

Diagonalizing the real-space excitation spectrum reveals a clear hierarchy of localized and extended magnon modes, shaped by the interplay between the hexameron cores and the surrounding frustrated $120^\circ$ magnetic background [Fig.~\ref{magnon_modes}(a)--\ref{magnon_modes}(c)]. The lowest excitation ($n = 1$ at $E_1 = 0.613~\mathrm{meV}$) is a hybridized quasi-Goldstone mode that retains finite amplitude across the $120^\circ$ background while showing sharp nodal modulations at the hexameron cores.
Above this mode, a discrete set of ten excitations ($n = 2 \dots 11$, spanning $E_n = 4.03$--$7.83~\mathrm{meV}$) appears cleanly inside a spin-wave gap below the bulk $120^\circ$ magnon continuum ($E \geq 9.18~\mathrm{meV}$), as shown in Fig.~\ref{magnon_modes}(b). Spatial probability density maps $|\psi_n(\mathbf{r}_i)|^2 = |u_{n,i}|^2 + |v_{n,i}|^2$ show that these in-gap states are tightly confined to the six-site perimeter of the hexameron cores, with negligible weight in the surrounding matrix [Fig.~\ref{magnon_modes}(c1)--\ref{magnon_modes}(c4)]. Each hexameron therefore acts as a real-space potential well that traps localized multipole magnon modes---including breathing ($m = 0$), dipolar ($m = \pm 1$), and quadrupolar ($m = \pm 2$) ring oscillations. More subtly, variations in the local exchange environment and scalar spin chirality across symmetry-related hexameron cores lift the exact degeneracy of their internal modes, splitting the lowest internal excitations into $E_2 = 4.026~\mathrm{meV}$ on Core~A and $E_3 = 4.304~\mathrm{meV}$ on Core~B, and providing a spectroscopic fingerprint that distinguishes the two cores.

Moreover, to characterize the momentum-space topology of the magnon bands, we calculated the gauge-invariant Berry curvature $\Omega_n(\mathbf{k})$ and the integer Chern numbers $C_n$ over the first Brillouin zone using the bosonic Fukui--Hatsugai--Suzuki plaquette formulation, adapted to the para-unitary metric $\eta = \mathrm{diag}(I_M, -I_M)$ (see Appendix~\ref{sm_fhs}~\cite{colpa1978diagonalization,shindou2013topological,fukui2005chern}):
\begin{equation}
    C_n = \frac{1}{2\pi} \int_{\mathrm{BZ}} d^2k \, \Omega_n(\mathbf{k}).
\end{equation}
Treating the hexameron supercell as the unit cell of a periodic hexameron array, Fig.~\ref{magnon_topology}(a) shows that the low-energy magnon branches are separated by finite gaps. Since the $120^\circ$ background is coplanar and carries no scalar chirality, these gaps are generated by the noncoplanar scalar spin chirality $\chi_{ijk} = \mathbf{S}_i \cdot (\mathbf{S}_j \times \mathbf{S}_k)$ of the hexameron texture and the broken inversion symmetry of the Janus lattice. Direct plaquette integration gives the quantized Chern numbers $C_1 = +1$, $C_3 = -1$, $C_4 = +1$, $C_7 = -2$, $C_8 = C_9 = +1$, and $C_{10} = C_{11} = -1$ (with $C_2 = C_5 = C_6 = C_{12} = 0$), satisfying the global sum rule $\sum_{n=1}^{M} C_n = 0$. The Berry curvature $\Omega_1(\mathbf{k})$ of the lowest branch shows a pronounced $C_3$-symmetric valley-contrasting pattern, concentrated sharply at the three alternating $K$ vertices of the hexagonal first Brillouin zone and vanishing near the $K'$ corners [Fig.~\ref{magnon_topology}(b)]. By the bosonic bulk-boundary correspondence, the non-vanishing winding numbers $W_1 = C_1 = +1$ above the lowest magnon branch and $W_{11} = \sum_{n=1}^{11} C_n = -1$ across the primary gap below the bulk continuum guarantee topologically protected chiral magnon edge states, as discussed in more detail in Supplementary Material Section V.

Finally, the nontrivial Berry curvature gives rise to a finite magnon thermal Hall response under a longitudinal temperature gradient. Within linear response theory, the two-dimensional sheet thermal Hall conductance is given by~\cite{matsumoto2011theoretical,murakami2017thermal}
\begin{equation}
    \kappa_{xy}^{2\mathrm{D}}(T) = -\frac{k_B^2 T}{\hbar} \sum_{n=1}^{M} \int_{\mathrm{BZ}} \frac{d^2k}{(2\pi)^2} \, c_2\!\left[g\big(\omega_n(\mathbf{k})\big)\right] \Omega_n(\mathbf{k}),
\end{equation}
where $g(\omega) = \left(e^{\hbar \omega / k_B T} - 1\right)^{-1}$ is the Bose--Einstein distribution and $c_2(x) = \int_0^x dt \, \left[\ln\left(\frac{1+t}{t}\right)\right]^2$. As shown in Fig.~\ref{magnon_topology}(c), $\kappa_{xy}^{2\mathrm{D}}(T)$ falls off exponentially as $T \to 0$ because thermal occupation is suppressed across the $0.613~\mathrm{meV}$ excitation gap. It then turns on smoothly and negatively above $T \approx 25~\mathrm{K}$ as the $C_1 = +1$ band becomes thermally populated, reaching $-1.281 \times 10^{-9}~\mathrm{W/K}$ at $T = 50~\mathrm{K}$. Importantly, the hexameron comes in two energetically degenerate partners related by time reversal~[see Fig.~S20], which flips all magnetic moments and therefore changes the sign of the scalar spin chirality. As a result, their magnon Berry curvatures satisfy \(\Omega_n^{(A)}(\mathbf{k})=-\Omega_{\bar n}^{(B)}(-\mathbf{k})\), where \(\bar n\) denotes the corresponding time-reversed magnon band. This leads to opposite Chern numbers and an inverted intrinsic thermal Hall response, \(\kappa_{xy}^{(B)}=-\kappa_{xy}^{(A)}\). In turn, the domain-dependent thermal Hall signal offers a way to tell the two partners apart in experiment, provided a hexameron array can be prepared. Experimentally, such a periodic lattice can be quenched into existence globally by rapidly sweeping an external magnetic field from the polarized or spin-flop regime, or via ultrafast optical and thermal quenches across the N\'eel temperature~\cite{jani2021antiferromagnetic,li2025intrinsic,juge2022skyrmions}. Conversely, isolated hexamerons or deterministic arrays can be nucleated by preparing the antiferromagnet in a uniform high-field state, rapidly quenching the field into the stability window of the target phase, and then applying a local writing stimulus to seed the first motif and guide replication across the array. In practice, one could field-cool the sample to the desired temperature, reduce the magnetic field in a fast step to the metastable regime where hexamerons are energetically accessible, and immediately address selected sites with nanosecond spin-orbit torque current pulses in heavy-metal/antiferromagnet heterostructures~\cite{amin2023antiferromagnetic,bhukta2026time}. Alternatively, voltage-controlled magnetic anisotropy gating can be used to transiently lower the local nucleation barrier before or during the quench, whereas tip-induced spin manipulation in spin-polarized scanning tunneling microscopy~\cite{romming2013writing,spethmann2020discovery} offers a site-specific route to write the initial seed and extend it deterministically into a patterned hexameron array.

\section*{CONCLUSIONS}
In summary, we have shown that the Janus MXene MnHf$_2$C$_2$FI realizes a frustrated triangular antiferromagnet in which metastable topological textures live on top of a $120^\circ$ N\'eel ground state. The asymmetric F/I termination confines magnetism to the Mn layer and breaks inversion symmetry, and our first-principles spin model reveals a dominant antiferromagnetic nearest-neighbor exchange together with a sizable Dzyaloshinskii--Moriya interaction. Besides the N\'eel ground state, this Hamiltonian supports the hexameron, a metastable composite of six merons and antimerons, one vortex--antivortex pair per sublattice, whose sublattice topological charges compensate. Under an out-of-plane field the hexameron reorganizes internally through sequential core-polarity reversals while keeping its winding structure, and it survives up to fields set by the large antiferromagnetic exchange. Its noncoplanar cores bind discrete in-gap magnon modes that are well separated from the bulk spin-wave continuum, and in a periodic hexameron array the magnon bands carry nonzero Chern numbers. The resulting magnon thermal Hall response changes sign between the two time-reversal partners of the hexameron and could therefore be used to distinguish them. Finally, the thermodynamics of the model shows an entropy deficit of $0.126\,k_{\mathrm B}$ per spin relative to a ferromagnetic reference, a quantitative measure of frustration in this classical magnet. Janus MXenes thus offer a two-dimensional platform in which frustration, broken inversion symmetry, and spin--orbit coupling combine to produce metastable antiferromagnetic solitons with clear spectroscopic and transport fingerprints.

\begin{acknowledgments}
\textbf{DPR} acknowledges the Paramrudra High-Performance Computing (HPC) facility at the Inter-University Accelerator Centre (IUAC), New Delhi, Govt. of India.\\
\textbf{SL} acknowledges funding by Deutsche Forschungsgemeinschaft (DFG): project 328545488--CRC/TRR 227, Project No. B12 and the German Excellence Strategy-EXC3112/1-533767171 (Center for Chiral Electronics).
\end{acknowledgments}

\appendix
\section{Triangular Lattice}
\label{triangular_spin}
For the study of this model system, the Ising Hamiltonian for a triangular lattice in the absence of a magnetic field is given by
\begin{align}
\mathbf{\hat{H} = J \sum_{<i,j>}[ \sigma_{i,\,j} \left( \sigma_{i+1,\,j} + \sigma_{i,\,j+1}\right) + \sigma_{i+1,\,j} \, \sigma_{i,\,j+1}]}
\end{align}

The Hamiltonian including the second-nearest-neighbor term is given by

\begin{align}
    \mathbf{\hat{H}} &= J_{1} \sum_{<i,j>}[ \sigma_{i,\,j} \left( \sigma_{i+1,\,j} + \sigma_{i,\,j+1}\right) + \sigma_{i+1,\,j} \, \sigma_{i,\,j+1} ] \nonumber \\
    &\quad + J_{2} \sum_{<i,j>}[ \sigma_{i,\,j} \left( \sigma_{i+1,\,j+1} + \sigma_{i+2,\,j-1}\right) + \sigma_{i+2,\,j-1} \, \sigma_{i+1,\,j+1} ] \label{eq:hamiltonian}
\end{align}

where the sum runs over all $(i,j)$ positions. Here, $J_{1}$ and $J_{2}$ are the coupling constants that determine the strength of the first- and next-nearest-neighbor interactions. For $J_1>0$ and $J_2\geq0$ are the antiferromagnetic nearest-neighbor and next-nearest-neighbor couplings.

For the honeycomb lattice,
\begin{equation}
H = J \sum_{\langle i,j\rangle} \left(1 - 2\,\delta_{ij,b}\right) \sigma_i \sigma_j ,
\end{equation}
where $\delta_{ij,b}=1$ on the ferromagnetic bonds $b$ and $0$ on the antiferromagnetic bonds. A hexagon with a single ferromagnetic bond cannot satisfy all of its bonds, every hexagon in a ground state has exactly one unsatisfied bond. These ground states are disordered and map onto the close-packed dimer model on the dual triangular lattice, with each unsatisfied bond acting as a dimer. 
The honeycomb lattice consists of $L\times L$ unit cells with two sites each (sublattices $A$ and $B$), 
so the total number of spins is $N = 2L^{2}$, with $3N/2$ nearest-neighbor bonds and $L^{2}$ hexagonal plaquettes. 
Periodic boundary conditions are applied in both lattice directions. 
The linear size $L$ must be even, because the ferromagnetic bonds pair neighboring hexagons so that each hexagon contains exactly one ferromagnetic bond.  
All thermodynamic quantities are reported per spin, i.e., divided by $N$.
\section{Calculation of Residual Entropy}
\label{Residual_Entropy}

For Ising spins, the procedure below yields the residual entropy; for classical vector spins, it yields the entropy deficit discussed in Sec.~\ref{sec:entropy}, and $k_B\ln 2^N$ is then replaced by the common high-temperature entropy of the two Hamiltonians.  
Integrating the thermodynamic relation gives the entropy of the system:
\begin{equation}
S(T_{f}) - S(0) = \int_{0}^{T_{f}} \frac{C_v}{T} \, dT
\end{equation}

\begin{equation}
S(T_{f}) = S(0) + \int_{0}^{T_{f}} \frac{C_v}{T} \, dT
\end{equation}

\begin{align}
S(T_{f})_{\text{ferro}} &= S(0)_{\text{ferro}} + \left(\int_{0}^{T_{f}} \frac{C_v}{T} \, dT \right)_{\text{ferro}} \nonumber \\
S(T_{f})_{\text{antiferro}} &= S(0)_{\text{antiferro}} + \left(\int_{0}^{T_{f}} \frac{C_v}{T} \, dT \right)_{\text{antiferro}}
\end{align}

At high temperature
\begin{eqnarray}
S(T_{f})_{\text{ferro}} = S(T_{f})_{\text{antiferro}} = k_{B}\ln 2^{N},
\end{eqnarray}
so that
\begin{eqnarray}
S(0)_{\text{ferro}} &+& \left(\int_{0}^{T_{f}} \frac{C_v}{T} dT\right)_{\text{ferro}} \nonumber \\
&=& S(0)_{\text{antiferro}}
+ \left(\int_{0}^{T_{f}} \frac{C_v}{T} dT\right)_{\text{antiferro}}
\end{eqnarray}

For the ferromagnetic case the system is ordered at low $T$, hence
\begin{eqnarray}
S(0)_{\text{ferro}} = 0 .
\end{eqnarray}
Therefore
\begin{eqnarray}
S(0)_{\text{antiferro}} &=& 
\left(\int_{0}^{T_{f}} \frac{C_v}{T} dT\right)_{\text{ferro}} \nonumber \\
&-& \left(\int_{0}^{T_{f}} \frac{C_v}{T} dT\right)_{\text{antiferro}}
\end{eqnarray}

Finally, the residual entropy is
\begin{eqnarray}
\boxed{ S(0)_{\text{antiferro}} =
\Delta S_{\text{ferro -minus antiferro}}^{\text{(high $T$)}} } .
\end{eqnarray}

\section{Atomistic spin Hamiltonian and magnetic dynamics}
\label{atomistic_model}

The magnetic exchange interactions were calculated based on the Wannier-interpolated
electronic structure. The exchange tensors obtained from the
noncollinear calculations were combined and subsequently used to construct the
atomistic spin model for simulations. The
effective magnetic Hamiltonian is expressed as
\begin{equation}
\label{H_total}
\mathcal{H}
=
\mathcal{H}_{\mathrm{ex}}
+
\mathcal{H}_{\mathrm{SIA}}
+
\mathcal{H}_{\mathrm{Z}}
+
\mathcal{H}_{\mathrm{dip}},
\end{equation}
where the terms represent exchange interactions, single-ion
anisotropy, the Zeeman interaction, and dipolar interactions,
respectively.


The exchange Hamiltonian
is written as
\begin{equation}
\label{H_exchange}
\mathcal{H}_{\mathrm{ex}}
=
-\sum_{i\ne j}
\mathbf{S}_{i}^{\mathrm{T}}
\boldsymbol{\mathcal{J}}_{ij}^{\mathrm{Full}}
\mathbf{S}_{j},
\end{equation}
where $\mathbf{S}_{i}=(S_i^x,S_i^y,S_i^z)^{\mathrm{T}}$
is the unit vector along the magnetic moment at site $i$,
and $\boldsymbol{\mathcal{J}}_{ij}^{\mathrm{Full}}$ is the
exchange tensor connecting sites $i$ and $j$.

The exchange tensor is decomposed into isotropic Heisenberg
exchange, antisymmetric exchange associated with the
Dzyaloshinskii--Moriya interaction (DMI), and symmetric
anisotropic exchange:
\begin{equation}
\label{J_decomposition}
\boldsymbol{\mathcal{J}}_{ij}^{\mathrm{Full}}
=
J_{ij}^{\mathrm{iso}}\mathbf{I}
+
\boldsymbol{\mathcal{J}}_{ij}^{\mathrm{DMI}}
+
\boldsymbol{\mathcal{J}}_{ij}^{\mathrm{ani}},
\end{equation}
where $J_{ij}^{\mathrm{iso}}$ is the isotropic exchange
parameter, $\mathbf{I}$ is the identity matrix,
$\boldsymbol{\mathcal{J}}_{ij}^{\mathrm{DMI}}$ is the
antisymmetric DMI tensor, and
$\boldsymbol{\mathcal{J}}_{ij}^{\mathrm{ani}}$ is the
symmetric anisotropic exchange tensor.

The isotropic Heisenberg contribution is
\begin{equation}
\label{eq:H_iso}
\mathcal{H}_{\mathrm{iso}}
=
-\sum_{i\ne j}
J_{ij}^{\mathrm{iso}}
\mathbf{S}_{i}\cdot\mathbf{S}_{j}.
\end{equation}
With this sign convention, positive
$J_{ij}^{\mathrm{iso}}$ favors parallel alignment, whereas
negative $J_{ij}^{\mathrm{iso}}$ favors antiparallel alignment.


The DMI contribution is
\begin{equation}
\label{eq:H_DMI}
\mathcal{H}_{\mathrm{DMI}}
=
-\sum_{i\ne j}
\mathbf{D}_{ij}\cdot
\left(\mathbf{S}_{i}\times\mathbf{S}_{j}\right),
\end{equation}
where $\mathbf{D}_{ij}=(D_{ij}^{x},D_{ij}^{y},D_{ij}^{z})$
is the DMI vector. The corresponding antisymmetric tensor
is defined as
\begin{equation}
\label{eq:J_DMI}
\boldsymbol{\mathcal{J}}_{ij}^{\mathrm{DMI}}
=
\begin{pmatrix}
0 & D_{ij}^{z} & -D_{ij}^{y}\\
-D_{ij}^{z} & 0 & D_{ij}^{x}\\
D_{ij}^{y} & -D_{ij}^{x} & 0
\end{pmatrix}.
\end{equation}
This interaction favors a particular sense of relative spin
rotation and can contribute to the formation of chiral
magnetic textures.


The symmetric anisotropic exchange contribution is
\begin{equation}
\label{eq:H_anisotropic_exchange}
\mathcal{H}_{\mathrm{ani-ex}}
=
-\sum_{i\ne j}
\mathbf{S}_{i}^{\mathrm{T}}
\boldsymbol{\mathcal{J}}_{ij}^{\mathrm{ani}}
\mathbf{S}_{j},
\end{equation}
where
\begin{equation}
\label{eq:J_ani_symmetric}
\boldsymbol{\mathcal{J}}_{ij}^{\mathrm{ani}}
=
\left(
\boldsymbol{\mathcal{J}}_{ij}^{\mathrm{ani}}
\right)^{\mathrm{T}}.
\end{equation}
This term accounts for the directional dependence of the
pairwise exchange interaction beyond the isotropic Heisenberg
model.

The combined exchange tensor can equivalently be written as
\begin{equation}
\label{eq:J_combined}
\boldsymbol{\mathcal{J}}_{ij}^{\mathrm{Full}}
=
\begin{pmatrix}
J_{xx} & J_{xy} & J_{xz}\\
J_{yx} & J_{yy} & J_{yz}\\
J_{zx} & J_{zy} & J_{zz}
\end{pmatrix}_{ij},
\end{equation}
where the diagonal elements contain the isotropic and
symmetric anisotropic contributions, while the antisymmetric
part of the off-diagonal elements contains the DMI.


When a uniaxial single-ion anisotropy is included, its
contribution is
\begin{equation}
\label{eq:H_SIA}
\mathcal{H}_{\mathrm{SIA}}
=
-\sum_i K_i
\left(\mathbf{S}_{i}\cdot\mathbf{e}_i\right)^2,
\end{equation}
where $K_i$ is the anisotropy energy and $\mathbf{e}_i$
is the unit vector along the anisotropy axis. This term
must be included only when the anisotropy parameter has
been determined and supplied to the simulation.


For simulations performed under an applied magnetic field,
the Zeeman contribution is
\begin{equation}
\label{eq:H_Zeeman}
\mathcal{H}_{\mathrm{Z}}
=
-\sum_i \mu_i
\mathbf{B}_{\mathrm{app}}\cdot\mathbf{S}_i,
\end{equation}
where $\mu_i$ is the magnitude of the magnetic moment at
site $i$ and $\mathbf{B}_{\mathrm{app}}$ is the applied
magnetic flux density.


If dipolar interactions are included, their contribution is
\begin{equation}
\label{eq:H_dip}
\mathcal{H}_{\mathrm{dip}}
=
\frac{\mu_0}{4\pi}
\sum_{i<j}
\frac{
\boldsymbol{\mu}_i\cdot\boldsymbol{\mu}_j
-
3(\boldsymbol{\mu}_i\cdot\hat{\mathbf{r}}_{ij})
 (\boldsymbol{\mu}_j\cdot\hat{\mathbf{r}}_{ij})
}{
r_{ij}^{3}
},
\end{equation}
where $\boldsymbol{\mu}_i=\mu_i\mathbf{S}_i$,
$\mathbf{r}_{ij}=\mathbf{r}_j-\mathbf{r}_i$,
$r_{ij}=|\mathbf{r}_{ij}|$, and
$\hat{\mathbf{r}}_{ij}=\mathbf{r}_{ij}/r_{ij}$.


The effective magnetic field acting on site $i$ is obtained
from the derivative of the total Hamiltonian:
\begin{equation}
\label{eq:B_eff}
\mathbf{B}_{i}^{\mathrm{eff}}
=
-\frac{1}{\mu_i}
\frac{\partial\mathcal{H}}{\partial\mathbf{S}_{i}},
\end{equation}
where the Hamiltonian is expressed in joules and $\mu_i$
in joules per tesla. The total effective field is the sum
of the contributions from the exchange, anisotropy,
applied-field, dipolar, and any other interactions
included in the model.


The time evolution of the normalized magnetic moments is
described by the Landau--Lifshitz--Gilbert (LLG) equation,
\begin{equation}
\label{eq:LLG2}
\frac{d\mathbf{S}_i}{dt}
=
-\frac{\gamma_i}{1+\alpha_i^2}
\left[
\mathbf{S}_i\times\mathbf{B}_i^{\mathrm{eff}}
+
\alpha_i\,
\mathbf{S}_i\times
\left(
\mathbf{S}_i\times\mathbf{B}_i^{\mathrm{eff}}
\right)
\right],
\end{equation}
where $\gamma_i$ is the gyromagnetic ratio and $\alpha_i$
is the dimensionless Gilbert damping parameter. The first
term describes precession around the effective field, while
the second term describes dissipative relaxation.

\section{Reciprocal-space exchange interactions and eigenvalue analysis}
\label{FourierJq}
The magnetic properties can be described by analyzing the Fourier-transformed magnetic interactions, which provide insight into the momentum-dependent exchange interactions and the energetics of spin-spiral configurations. The real-space exchange tensors were then transformed into reciprocal space using the Fourier convention

\begin{equation}
\mathcal{J}_{ij}^{\alpha\beta}(\mathbf q)
=
-\sum_{\mathbf R}
H_{ij}^{\alpha\beta}(\mathbf R)
e^{2\pi i\mathbf q\cdot\mathbf R},
\label{Jq_fourier}
\end{equation}
where $i,j$ label magnetic sites, $\alpha,\beta\in\{x,y,z\}$ denote Cartesian spin components, $\mathbf R$ is a lattice translation vector, and $\mathbf q$ is expressed in fractional reciprocal-lattice coordinates. The real-space interaction Hessian is constructed as

\begin{equation}
H_{ij}(\mathbf R)
=
J_{ij}^{\mathrm{iso}}(\mathbf R)\mathbb I_3
+
H_{ij}^{\mathrm{DMI}}(\mathbf R)
+
\Gamma_{ij}(\mathbf R),
\label{exchange_hessian}
\end{equation}
where $J_{ij}^{\mathrm{iso}}$ denotes the isotropic Heisenberg exchange, $\Gamma_{ij}$ represents the symmetric anisotropic exchange tensor, and the DMI contribution is expressed as

\begin{equation}
H_{ij}^{\mathrm{DMI}}(\mathbf R)=
\begin{pmatrix}
0 & D_{ij}^{z} & -D_{ij}^{y}\\
-D_{ij}^{z} & 0 & D_{ij}^{x}\\
D_{ij}^{y} & -D_{ij}^{x} & 0
\end{pmatrix}.
\label{dmi_hessian}
\end{equation}

The resulting momentum-dependent interaction matrix was diagonalized by solving

\begin{equation}
\mathcal J(\mathbf q)\mathbf v_n(\mathbf q)
=
\lambda_n(\mathbf q)\mathbf v_n(\mathbf q),
\label{Jq_eigenvalues}
\end{equation}
where $\lambda_n(\mathbf q)$ and $\mathbf v_n(\mathbf q)$ denote the corresponding eigenvalues and normalized eigenvectors, respectively. Examining the eigenvalue dispersion together with the corresponding eigenvector polarizations provides insight into the momentum-dependent magnetic interactions and the preferred spin-polarization modes.

\section{Bosonic Fukui--Hatsugai--Suzuki Formulation for Magnon Chern Numbers}
\label{sm_fhs}

Bosonic Bogoliubov quasiparticles obey canonical commutation relations rather than fermionic anticommutation relations. The quadratic spin-wave Hamiltonian $\mathcal{H}_{\mathbf{k}}$ is diagonalized via a para-unitary transformation $T_{\mathbf{k}}$ satisfying
\begin{equation}
T_{\mathbf{k}}^\dagger \eta T_{\mathbf{k}} = \eta,
\qquad
\eta =
\begin{pmatrix}
I_M & 0 \\
0 & -I_M
\end{pmatrix},
\end{equation}
such that
\begin{equation}
T_{\mathbf{k}}^\dagger
\mathcal{H}_{\mathbf{k}}
T_{\mathbf{k}}
=
\mathrm{diag}
\left[
\omega_1(\mathbf{k}), \dots, \omega_M(\mathbf{k}),
\omega_1(-\mathbf{k}), \dots, \omega_M(-\mathbf{k})
\right],
\end{equation}
where $M$ denotes the number of magnetic sublattices in the supercell
\cite{colpa1978diagonalization,shindou2013topological}.

Selecting the $M$ positive-energy particle eigenvectors
$|\psi_n(\mathbf{k})\rangle$, normalized according to the indefinite bosonic inner product
\begin{equation}
\langle \psi_n(\mathbf{k}) |
\eta |
\psi_m(\mathbf{k}) \rangle
=
\delta_{nm},
\end{equation}
the gauge-invariant Berry curvature is evaluated on a discretized $Nk = 50$, i.e
$50 \times 50$ Brillouin-zone mesh using the bosonic adaptation
of the Fukui--Hatsugai--Suzuki lattice method
\cite{fukui2005chern}.

For each isolated magnon branch $n \in \{1,\dots,M\}$,
we define the bosonic $U(1)$ link variable along the primitive
reciprocal-lattice displacement $\delta\mathbf{k}_{\mu}$
($\mu=1,2$) as
\begin{equation}
U_{\mu}^{(n)}(\mathbf{k})
=
\frac{
\langle \psi_n(\mathbf{k}) |
\eta |
\psi_n(\mathbf{k}+\delta\mathbf{k}_{\mu}) \rangle
}{
\left|
\langle \psi_n(\mathbf{k}) |
\eta |
\psi_n(\mathbf{k}+\delta\mathbf{k}_{\mu}) \rangle
\right|
}.
\end{equation}

The gauge-invariant lattice Berry flux
$F_{12}^{(n)}(\mathbf{k})$ piercing an elementary momentum-space
plaquette is then obtained from the closed Wilson loop around
the plaquette vertices:
\begin{equation}
\begin{aligned}
F_{12}^{(n)}(\mathbf{k})
=
\operatorname{Im}\ln \Big[
&U_1^{(n)}(\mathbf{k})
U_2^{(n)}(\mathbf{k}+\delta\mathbf{k}_1)\\
&\times
U_1^{(n)}(\mathbf{k}+\delta\mathbf{k}_2)^{*}
U_2^{(n)}(\mathbf{k})^{*}
\Big].
\end{aligned}
\end{equation}
The flux is restricted to the principal branch
$-\pi < F_{12}^{(n)}(\mathbf{k}) \leq \pi$.

Dividing $F_{12}^{(n)}(\mathbf{k})$ by the plaquette area
$\Delta A_{\mathrm{BZ}}=A_{\mathrm{BZ}}/N_k^2$ yields the
local Berry curvature $\Omega_n(\mathbf{k})$, while summing
over all $N_k^2$ plaquettes in the first Brillouin zone
produces the quantized integer Chern number:
\begin{equation}
C_n
=
\frac{1}{2\pi}
\sum_{\mathbf{k}}
F_{12}^{(n)}(\mathbf{k})
\in \mathbb{Z}.
\end{equation}


\bibliography{apssamp}

@article{balents2010spin,
  title={Spin liquids in frustrated magnets},
  author={Balents, Leon},
  journal={nature},
  volume={464},
  number={7286},
  pages={199--208},
  year={2010},
  publisher={Nature Publishing Group UK, London}
}

@article{buessen2018quantum,
  title={Quantum spin liquids in frustrated spin-1 diamond antiferromagnets},
  author={Buessen, Finn Lasse and Hering, Max and Reuther, Johannes and Trebst, Simon},
  journal={Physical review letters},
  volume={120},
  number={5},
  pages={057201},
  year={2018},
  publisher={APS}
}

@article{chatterjee2026semi,
  title={Semi-Dirac spin liquids and frustrated quantum magnetism on the trellis lattice},
  author={Chatterjee, Sourin and Maity, Atanu and Potten, Janik and M{\"u}ller, Tobias and Feuerpfeil, Andreas and Thomale, Ronny and Penc, Karlo and Jeschke, Harald O and Samajdar, Rhine and Iqbal, Yasir},
  journal={Physical Review Research},
  volume={8},
  number={1},
  pages={013191},
  year={2026},
  publisher={APS}
}

@article{jena2026spin,
  title={Spin liquid state in the three dimensional pyrochlore-like frustrated magnet MgCrGaO 4},
  author={Jena, U and Kundu, S and Lee, Suheon and Faure, Q and Damay, F and Rols, S and Berlie, Adam and Petit, S and Choi, Kwang-Yong and Khuntia, P},
  journal={Physical Review B},
  volume={113},
  number={10},
  pages={L100404},
  year={2026},
  publisher={APS}
}

@article{li2026kitaev,
  title={Kitaev-derived spin liquid in the frustrated quantum magnet Na2Co2TeO6},
  author={Li, Han and Zhou, Xu-Guang and Su, Gang and Li, Wei},
  journal={npj Quantum Materials},
  year={2026},
  publisher={Nature Publishing Group UK, London}
}

@article{ma2026coexistence,
  title={Coexistence of mixed-dimensional fluctuations and incommensurate long-range order in the frustrated antiferromagnet C e 5 Ag Pb 3},
  author={Ma, Zhijie and Sun, Ying and Luo, Jiechen and Deng, Sihao and Wang, Hongde and Yuan, Xiuliang and Zhang, Kaiqi and He, Lunhua and Wang, Cong and Shi, Youguo},
  journal={Physical Review B},
  volume={114},
  number={2},
  pages={024406},
  year={2026},
  publisher={APS}
}

@article{cheng2026interwoven,
  title={Interwoven magnetic kagome metal overcomes geometric frustration},
  author={Cheng, Erjian and Wang, Kaipu and Hao, Yiqing and Chen, Wenqing and Tan, Hengxin and Li, Zongkai and Wang, Meixiao and Gao, Wenli and Wu, Di and Sun, Shuaishuai and others},
  journal={Nature materials},
  volume={25},
  number={4},
  pages={602--609},
  year={2026},
  publisher={Nature Publishing Group UK, London}
}

@article{PhysRev.79.357,
  title = {Antiferromagnetism. The Triangular Ising Net},
  author = {Wannier, G. H.},
  journal = {Phys. Rev.},
  volume = {79},
  issue = {2},
  pages = {357--364},
  numpages = {0},
  year = {1950},
  month = {Jul},
  publisher = {American Physical Society},
  doi = {10.1103/PhysRev.79.357},
  url = {https://link.aps.org/doi/10.1103/PhysRev.79.357}
}

@article{zhou2025triple,
  title={Triple-Q state in magnetic breathing kagome lattice},
  author={Zhou, Hangyu and dos Santos Dias, Manuel and Bao, Shijian and Lu, Hanchen and Zhang, Youguang and Zhao, Weisheng and Lounis, Samir},
  journal={npj Spintronics},
  volume={3},
  number={1},
  pages={31},
  year={2025},
  publisher={Nature Publishing Group UK London}
}

@book{diep2013frustrated,
  title={Frustrated spin systems},
  author={Diep, Hung T and others},
  year={2013},
  publisher={World scientific}
}

@article{ramirez1994strongly,
  title={Strongly geometrically frustrated magnets},
  author={Ramirez, AP},
  journal={Annual Review of Materials Science},
  volume={24},
  number={1},
  pages={453--480},
  year={1994},
  publisher={Annual Reviews 4139 El Camino Way, PO Box 10139, Palo Alto, CA 94303-0139, USA}
}

@article{villain1979insulating,
  title={Insulating spin glasses},
  author={Villain, Jacques},
  journal={Zeitschrift f{\"u}r Physik B Condensed Matter},
  volume={33},
  number={1},
  pages={31--42},
  year={1979},
  publisher={Springer}
}

@article{laforge2013quasispin,
  title={Quasispin glass in a geometrically frustrated magnet},
  author={LaForge, Andrew D and Pulido, Sergio H and Cava, Robert Joseph and Chan, Benny C and Ramirez, Arthur P},
  journal={Physical review letters},
  volume={110},
  number={1},
  pages={017203},
  year={2013},
  publisher={APS}
}

@article{jena2026topological,
  title={Topological spin freezing in frustrated quantum materials},
  author={Jena, U and Barman, M and Pradhan, A and Khuntia, P},
  journal={arXiv preprint arXiv:2605.06069},
  year={2026}
}

@article{hromov2025incommensurate,
  title={Incommensurate magnetic order arising from frustrated interchain interactions in the spin-1 2 chain compound AgCuVO 4},
  author={Hromov, Anton and Zorko, Andrej and Gomil{\v{s}}ek, Matja{\v{z}} and Orench, I Puente and Keller, Lukas and Shiroka, Toni and Prokofiev, Andrey and Pregelj, Matej},
  journal={Physical Review B},
  volume={112},
  number={13},
  pages={134433},
  year={2025},
  publisher={APS}
}

@article{chung2005statics,
  title={Statics and dynamics of incommensurate spin order in a geometrically frustrated antiferromagnet CdCr 2 O 4},
  author={Chung, J-H and Matsuda, M and Lee, S-H and Kakurai, K and Ueda, H and Sato, TJ and Takagi, H and Hong, K-P and Park, S},
  journal={Physical review letters},
  volume={95},
  number={24},
  pages={247204},
  year={2005},
  publisher={APS}
}

@article{aldarawsheh2023intrinsic,
  title={Intrinsic N{\'e}el antiferromagnetic multimeronic spin textures in ultrathin films},
  author={Aldarawsheh, Amal and Sallermann, Moritz and Abusaa, Muayad and Lounis, Samir},
  journal={The journal of physical chemistry letters},
  volume={14},
  number={40},
  pages={8970},
  year={2023}
}

@article{mostofi2008wannier90,
  title={wannier90: A tool for obtaining maximally-localised Wannier functions},
  author={Mostofi, Arash A and Yates, Jonathan R and Lee, Young-Su and Souza, Ivo and Vanderbilt, David and Marzari, Nicola},
  journal={Computer physics communications},
  volume={178},
  number={9},
  pages={685--699},
  year={2008},
  publisher={Elsevier}
}

@article{kresse1996efficient,
  title={Efficient iterative schemes for ab initio total-energy calculations using a plane-wave basis set},
  author={Kresse, Georg and Furthm{\"u}ller, J{\"u}rgen},
  journal={Physical review B},
  volume={54},
  number={16},
  pages={11169},
  year={1996},
  publisher={APS}
}

@article{kresse1993ab,
  title={Ab initio molecular dynamics for liquid metals},
  author={Kresse, Georg and Hafner, J{\"u}rgen},
  journal={Physical review B},
  volume={47},
  number={1},
  pages={558},
  year={1993},
  publisher={APS}
}

@article{blochl1994projector,
  title={Projector augmented-wave method},
  author={Bl{\"o}chl, Peter E},
  journal={Physical review B},
  volume={50},
  number={24},
  pages={17953},
  year={1994},
  publisher={APS}
}

@article{liechtenstein1995density,
  title={Density-functional theory and strong interactions: Orbital ordering in Mott-Hubbard insulators},
  author={Liechtenstein, AI and Anisimov, Vladimir I and Zaanen, Jan},
  journal={Physical Review B},
  volume={52},
  number={8},
  pages={R5467},
  year={1995},
  publisher={APS}
}

@article{dudarev1998electron,
  title={Electron-energy-loss spectra and the structural stability of nickel oxide: An LSDA+ U study},
  author={Dudarev, Sergei L and Botton, Gianluigi A and Savrasov, Sergey Y and Humphreys, Colin J and Sutton, Adrian P},
  journal={Physical Review B},
  volume={57},
  number={3},
  pages={1505},
  year={1998},
  publisher={APS}
}

@article{perdew1996generalized,
  title={Generalized gradient approximation made simple},
  author={Perdew, John P and Burke, Kieron and Ernzerhof, Matthias},
  journal={Physical review letters},
  volume={77},
  number={18},
  pages={3865},
  year={1996},
  publisher={APS}
}

@article{grimme2010consistent,
  title={A consistent and accurate ab initio parametrization of density functional dispersion correction (DFT-D) for the 94 elements H-Pu},
  author={Grimme, Stefan and Antony, Jens and Ehrlich, Stephan and Krieg, Helge},
  journal={The Journal of chemical physics},
  volume={132},
  number={15},
  year={2010},
  publisher={AIP Publishing}
}

@article{wannier1973antiferromagnetism,
  title={Antiferromagnetism. the triangular ising net},
  author={Wannier, GH},
  journal={Physical Review B},
  volume={7},
  number={11},
  pages={5017},
  year={1973},
  publisher={APS}
}

@article{misguich1999spin,
  title={Spin-liquid phase of the multiple-spin exchange Hamiltonian on the triangular lattice},
  author={Misguich, Gregoire and Lhuillier, Claire and Bernu, Bernard and Waldtmann, Christian},
  journal={Physical Review B},
  volume={60},
  number={2},
  pages={1064},
  year={1999},
  publisher={APS}
}

@article{abuawwad2022noncollinear,
  title={Noncollinear magnetism in two-dimensional CrTe2},
  author={Abuawwad, Nihad and dos Santos Dias, Manuel and Abusara, Hazem and Lounis, Samir},
  journal={Journal of Physics: Condensed Matter},
  volume={34},
  number={45},
  pages={454001},
  year={2022},
  publisher={IOP Publishing}
}

@article{nagaosa2013topological,
  title={Topological properties and dynamics of magnetic skyrmions},
  author={Nagaosa, Naoto and Tokura, Yoshinori},
  journal={Nature nanotechnology},
  volume={8},
  number={12},
  pages={899--911},
  year={2013},
  publisher={Nature Publishing Group}
}

@article{souza2026dynamics,
  title={Dynamics and pinning for skyrmions in altermagnets},
  author={Souza, JC Bellizotti and Reichhardt, CJO and Saxena, A and Reichhardt, C},
  journal={Physical Review B},
  volume={113},
  number={22},
  pages={224424},
  year={2026},
  publisher={APS}
}

@article{yuan2020intrinsic,
  title={Intrinsic skyrmions in monolayer Janus magnets},
  author={Yuan, Jiaren and Yang, Yumeng and Cai, Yongqing and Wu, Yihong and Chen, Yuanping and Yan, Xiaohong and Shen, Lei},
  journal={Physical Review B},
  volume={101},
  number={9},
  pages={094420},
  year={2020},
  publisher={APS}
}

@article{nayak2017magnetic,
  title={Magnetic antiskyrmions above room temperature in tetragonal Heusler materials},
  author={Nayak, Ajaya K and Kumar, Vivek and Ma, Tianping and Werner, Peter and Pippel, Eckhard and Sahoo, Roshnee and Damay, Franoise and R{\"o}{\ss}ler, Ulrich K and Felser, Claudia and Parkin, Stuart SP},
  journal={Nature},
  volume={548},
  number={7669},
  pages={561--566},
  year={2017},
  publisher={Nature Publishing Group UK London}
}

@article{amin2023antiferromagnetic,
  title={Antiferromagnetic half-skyrmions electrically generated and controlled at room temperature},
  author={Amin, OJ and Poole, SF and Reimers, S and Barton, LX and Dal Din, A and Maccherozzi, F and Dhesi, SS and Nov{\'a}k, V{\'\i}t and Krizek, F and Chauhan, JS and others},
  journal={Nature Nanotechnology},
  volume={18},
  number={8},
  pages={849--853},
  year={2023},
  publisher={Nature Publishing Group UK London}
}

@article{li2026hybrid,
  title={Hybrid magnetic skyrmions with near-zero Hall angle and electrical switchability in a 2D multiferroic},
  author={Li, Xuhong and Zhou, Mo and Wei, Yongteng and Cao, Tengfei and Fan, Xiaoli},
  journal={npj Computational Materials},
  volume={12},
  number={1},
  pages={148},
  year={2026},
  publisher={Nature Publishing Group UK London}
}

@article{bhukta2024homochiral,
  title={Homochiral antiferromagnetic merons, antimerons and bimerons realized in synthetic antiferromagnets},
  author={Bhukta, Mona and Dohi, Takaaki and Bharadwaj, Venkata Krishna and Zarzuela, Ricardo and Syskaki, Maria-Andromachi and Foerster, Michael and Ni{\~n}o, Miguel Angel and Sinova, Jairo and Fr{\"o}mter, Robert and Kl{\"a}ui, Mathias},
  journal={Nature Communications},
  volume={15},
  number={1},
  pages={1641},
  year={2024},
  publisher={Nature Publishing Group UK London}
}

@article{chaix2026triangular,
  title={From triangular correlated paramagnet to multi-q noncoplanar spin state in spinel GeFe 2 O 4},
  author={Chaix, L and Robert, J and Chan, E and Ressouche, E and Petit, S and Colin, CV and Ballou, R and Ollivier, J and Regnault, L-P and Lhotel, E and others},
  journal={Physical Review Letters},
  volume={136},
  number={1},
  pages={016703},
  year={2026},
  publisher={APS}
}

@article{okamura2026giant,
  title={Giant topological magneto-optical effect in noncoplanar antiferromagnet},
  author={Okamura, Y and Hayashi, Y and Khanh, ND and Tokura, Y and Seki, S and Takahashi, Y},
  journal={Nature Communications},
  volume={17},
  number={1},
  pages={4409},
  year={2026},
  publisher={Nature Publishing Group UK London}
}

@article{sozeri2026frustrated,
  title={Frustrated magnetism in Mn films on Ag (111) surface: From chiral in-plane N{\'e}el state to row-wise antiferromagnetism},
  author={S{\"o}zeri, Selcuk and Abuawwad, Nihad and Aldarawsheh, Amal and Lounis, Samir},
  journal={Physical Review B},
  volume={113},
  number={10},
  pages={104415},
  year={2026},
  publisher={APS}
}

@article{gao2008revealing,
  title={Revealing the 120 antiferromagnetic N{\'e}el structure in real space: One monolayer Mn on Ag (111)},
  author={Gao, CL and Wulfhekel, W and Kirschner, J},
  journal={Physical Review Letters},
  volume={101},
  number={26},
  pages={267205},
  year={2008},
  publisher={APS}
}

@article{rodriguez2024phase,
  title={Phase coexistence of Mn trimer clusters and antiferromagnetic Mn islands on Ir (111)},
  author={Rodr{\'\i}guez-Sota, Arturo and Saxena, Vishesh and Spethmann, Jonas and Wiesendanger, Roland and Lo Conte, Roberto and Kubetzka, Andr{\'e} and von Bergmann, Kirsten},
  journal={ACS nano},
  volume={18},
  number={4},
  pages={3699--3706},
  year={2024},
  publisher={ACS Publications}
}

@article{kameda2026interactions,
  title={Interactions between atomic-scale skyrmions in 2D chiral magnets},
  author={Kameda, Mai and Kobayashi, Koji and Kawaguchi, Yuki},
  journal={Scientific Reports},
  volume={16},
  number={1},
  pages={12941},
  year={2026},
  publisher={Nature Publishing Group UK London}
}

@article{birch2022history,
  title={History-dependent domain and skyrmion formation in 2D van der Waals magnet Fe3GeTe2},
  author={Birch, MT and Powalla, L and Wintz, S and Hovorka, Ondrej and Litzius, Kai and Loudon, JC and Turnbull, LA and Nehruji, Vanessa and Son, K and Bubeck, C and others},
  journal={Nature communications},
  volume={13},
  number={1},
  pages={3035},
  year={2022},
  publisher={Nature Publishing Group UK London}
}

@article{heinze2011spontaneous,
  title={Spontaneous atomic-scale magnetic skyrmion lattice in two dimensions},
  author={Heinze, Stefan and Von Bergmann, Kirsten and Menzel, Matthias and Brede, Jens and Kubetzka, Andr{\'e} and Wiesendanger, Roland and Bihlmayer, Gustav and Bl{\"u}gel, Stefan},
  journal={nature physics},
  volume={7},
  number={9},
  pages={713--718},
  year={2011},
  publisher={Nature Publishing Group UK London}
}

@article{kabiraj2023realizing,
  title={Realizing unipolar and bipolar intrinsic skyrmions in MXenes from high-fidelity first-principles calculations},
  author={Kabiraj, Arnab and Mahapatra, Santanu},
  journal={NPJ Computational Materials},
  volume={9},
  number={1},
  pages={173},
  year={2023},
  publisher={Nature Publishing Group UK London}
}

@article{abuawwad2024electrical,
  title={Electrical engineering of topological magnetism in two-dimensional heterobilayers},
  author={Abuawwad, Nihad and dos Santos Dias, Manuel and Abusara, Hazem and Lounis, Samir},
  journal={npj Spintronics},
  volume={2},
  number={1},
  pages={10},
  year={2024},
  publisher={Nature Publishing Group UK London}
}

@article{mokkath2025field,
  title={Field-tunable skyrmion phases in monolayer MXene for spintronic applications},
  author={Mokkath, Junais Habeeb},
  journal={Physical Chemistry Chemical Physics},
  volume={27},
  number={35},
  pages={18152--18161},
  year={2025},
  publisher={The Royal Society of Chemistry}
}

@article{naguib2011two,
  title={Two-dimensional nanocrystals: two-dimensional nanocrystals produced by exfoliation of Ti3AlC2 (Adv. Mater. 37/2011)},
  author={Naguib, Michael and Kurtoglu, Murat and Presser, Volker and Lu, Jun and Niu, Junjie and Heon, Min and Hultman, Lars and Gogotsi, Yury and Barsoum, Michel W},
  journal={Advanced Materials},
  volume={23},
  number={37},
  pages={4207--4207},
  year={2011},
  publisher={Wiley Online Library}
}

@article{subba2025comprehensive,
  title={Comprehensive review of MAX phase and MXene materials: synthesis, properties, and applications},
  author={Subba, C Bahadur and Rai, Dibya Prakash and Tursunov, Mukhriddin E and Dekhkonov, Avazbek T and Pachuau, Z},
  journal={Carbon Letters},
  volume={35},
  number={6},
  pages={2485--2570},
  year={2025},
  publisher={Springer}
}

@article{kumar2022methods,
  title={Methods of synthesis, characteristics, and environmental applications of MXene: A comprehensive review},
  author={Kumar, Jagadeesan Aravind and Prakash, Pandurangan and Krithiga, Thangavelu and Amarnath, Duvuru Joshua and Premkumar, Jayapal and Rajamohan, Natarajan and Vasseghian, Yasser and Saravanan, Panchamoorthy and Rajasimman, Manivasagan},
  journal={Chemosphere},
  volume={286},
  pages={131607},
  year={2022},
  publisher={Elsevier}
}

@article{ajmal2025advancements,
  title={Advancements in MXene-based frameworks towards photocatalytic hydrogen production, carbon dioxide reduction and pollutant degradation: Current challenges and future prospects},
  author={Ajmal, Zeeshan and Hayat, Asif and Qadeer, Abdul and Zhao, Yu and Ibrahim, Essam H and ul Haq, Mahmood and Iqbal, Kanwal and Imran, Mohd and Kuku, Mohammed and Hussain, Iftikhar and others},
  journal={Coordination Chemistry Reviews},
  volume={523},
  pages={216226},
  year={2025},
  publisher={Elsevier}
}

@article{subba20252d,
  author    = {C. B. Subba and Prasad Mattipally and J. Sivakumar and A. Laref and A. Yvaz and D. P. Rai and Z. Pachuau},
  title     = {2D MXene-based photocatalysts for efficient water splitting and hydrogen evolution: a brief review},
  journal   = {Journal of the Iranian Chemical Society},
  year      = {2025},
  volume    = {22},
  number    = {9},
  pages     = {1825--1848},
  doi       = {10.1007/s13738-025-03273-1},
  url       = {https://doi.org/10.1007/s13738-025-03273-1},
  issn      = {1735-2428}
}

@article{nag2025modulation,
  title={Modulation of the Electronic Structure of Ti3C2T x MXene via Thiol Functionalization Induced by a Click Reaction},
  author={Nag, Riya and Chakraborty, Raja and Bar, Rajshekhar and Layek, Rama Kanta and Bera, Abhijit},
  journal={The Journal of Physical Chemistry C},
  volume = {129},
  number = {25},
  pages = {11476-11486},
  year={2025},
  publisher={ACS Publications}
}

@article{MOKKATH2025112967,
title = {Crucial influence of surface terminations on the ,electronic, optical, and vibrational properties of Nb2CTx MXene},
journal = {Materials Today Communications},
volume = {47},
pages = {112967},
year = {2025},
issn = {2352-4928},
doi = {https://doi.org/10.1016/j.mtcomm.2025.112967},
url = {https://www.sciencedirect.com/science/article/pii/S2352492825014795},
author = {Junais Habeeb Mokkath}
}

@Article{D5CP01402F,
author ="Bu, Yulin and Sun, Mengtao",
title  ="Tuning the electronic properties of WS2/Sc2C heterostructures via surface functionalization: a first-principles study",
journal  ="Phys. Chem. Chem. Phys.",
year  ="2025",
volume  ="27",
issue  ="27",
pages  ="14397-14409",
publisher  ="The Royal Society of Chemistry",
doi  ="10.1039/D5CP01402F",
url  ="http://dx.doi.org/10.1039/D5CP01402F"}

@article{doi:10.1021/acs.jpclett.0c03710,
author = {Ibragimova, Rina and Erhart, Paul and Rinke, Patrick and Komsa, Hannu-Pekka},
title = {Surface Functionalization of 2D MXenes: Trends in Distribution, Composition, and Electronic Properties},
journal = {The Journal of Physical Chemistry Letters},
volume = {12},
number = {9},
pages = {2377-2384},
year = {2021},
doi = {10.1021/acs.jpclett.0c03710},
    note ={PMID: 33657317},

URL = { 
    
        https://doi.org/10.1021/acs.jpclett.0c03710
    
    

},
eprint = { 
    
        https://doi.org/10.1021/acs.jpclett.0c03710
    
    

}

}

@incollection{prakash2025mxene,
  title={MXene-Based Materials for Photocatalytic Water Splitting},
  author={Prakash, R Mithun and Sakar, M},
  booktitle={MXenes: Expanding the Frontiers of Energy Applications},
  pages={155--188},
  year={2025},
  publisher={Springer}
}

@article{mahapatra2025artificial,
  title={Artificial intelligence interventions in 2D MXenes-based photocatalytic applications},
  author={Mahapatra, Durga Madhab and Kumar, Ashish and Kumar, Rajesh and Gupta, Navneet Kumar and Ethiraj, Baranitharan and Singh, Lakhveer},
  journal={Coordination Chemistry Reviews},
  volume={529},
  pages={216460},
  year={2025},
  publisher={Elsevier}
}

@article{stokes2005findsym,
  title={FINDSYM: program for identifying the space-group symmetry of a crystal},
  author={Stokes, Harold T and Hatch, Dorian M},
  journal={Applied Crystallography},
  volume={38},
  number={1},
  pages={237--238},
  year={2005},
  publisher={International Union of Crystallography}
}

@article{moore2024high,
  title={High-throughput determination of Hubbard U and Hund J values for transition metal oxides via the linear response formalism},
  author={Moore, Guy C and Horton, Matthew K and Linscott, Edward and Ganose, Alexander M and Siron, Martin and O'Regan, David D and Persson, Kristin A},
  journal={Physical Review Materials},
  volume={8},
  number={1},
  pages={014409},
  year={2024},
  publisher={APS}
}

@article{wang2006oxidation,
  title={Oxidation energies of transition metal oxides within the GGA+ U framework},
  author={Wang, Lei and Maxisch, Thomas and Ceder, Gerbrand},
  journal={Physical Review B—Condensed Matter and Materials Physics},
  volume={73},
  number={19},
  pages={195107},
  year={2006},
  publisher={APS}
}

@article{jain2011formation,
  title={Formation enthalpies by mixing GGA and GGA+ U calculations},
  author={Jain, Anubhav and Hautier, Geoffroy and Ong, Shyue Ping and Moore, Charles J and Fischer, Christopher C and Persson, Kristin A and Ceder, Gerbrand},
  journal={Physical Review B—Condensed Matter and Materials Physics},
  volume={84},
  number={4},
  pages={045115},
  year={2011},
  publisher={APS}
}

@article{spiritPaper,
  title = {Spirit: Multifunctional framework for atomistic spin simulations},
  author = {M\"uller, Gideon P. and Hoffmann, Markus and Di\ss{}elkamp, Constantin and Sch\"urhoff, Daniel and Mavros, Stefanos and Sallermann, Moritz and Kiselev, Nikolai S. and J\'onsson, Hannes and Bl\"ugel, Stefan},
  journal = {Phys. Rev. B},
  volume = {99},
  issue = {22},
  pages = {224414},
  numpages = {16},
  year = {2019},
  month = {Jun},
  publisher = {American Physical Society},
  doi = {10.1103/PhysRevB.99.224414},
  url = {https://link.aps.org/doi/10.1103/PhysRevB.99.224414}
}

@misc{spiritCode,
  author       = {Müller, Gideon P. and Sallermann, Moritz and Mavros, Stefanos and Rhiem, Florian and Schürhoff, Daniel and Hoffmann, Markus and Meyer, Ingo and Disselkamp, Constantin and Redies, Matthias and Buhl, Patrick and Suckert, Jens Rene and Ivanov, Aleksei V. and Kiselev, Nikolai S. and Jónsson, Hannes and Blügel, Stefan},
  title        = {{SPIRIT}},
  month        = mar,
  year         = 2023,
  publisher    = {Zenodo},
  doi          = {10.5281/zenodo.7746552},
  url          = {https://doi.org/10.5281/zenodo.7746552},
  howpublished  = {Zenodo}
}

@article{seki2012formation,
  title={Formation and rotation of skyrmion crystal in the chiral-lattice insulator Cu 2 OSeO 3},
  author={Seki, S and Kim, J-H and Inosov, DS and Georgii, R and Keimer, B and Ishiwata, S and Tokura, Y},
  journal={Physical Review B—Condensed Matter and Materials Physics},
  volume={85},
  number={22},
  pages={220406},
  year={2012},
  publisher={APS}
}

@article{lee2026asymmetrically,
  title={An asymmetrically out-of-plane ordered MAX phase as a precursor for Janus MXenes},
  author={Lee, Minseok and Seong, Hyun Woo and Ryu, Ho Jin},
  journal={Nature Communications},
  volume={17},
  pages={5910},
  year={2026}
}

@article{helmi2025novel,
  title={Novel Method for Synthesis of Janus MXene Nanosheets by Masking Starch},
  author={Helmi, Fatemeh and Yavari, Mahdieh and Yaghoubi-Azbari, Esmail and Akbari, Mahdi and Shariaty-Niassar, Mojtaba},
  journal={International Journal of Nanoscience and Nanotechnology},
  volume={21},
  number={3},
  pages={149--156},
  year={2025},
  publisher={Iranian Nanotechnology Society}
}

@article{thasitha2025janus,
  title={Janus MoWN MXene with tunable functional groups: structural stability, mechanical property, and phonon-mediated superconductivity},
  author={Thasitha, Sirinee and Bovornratanaraks, Thiti and Kotmool, Komsilp},
  journal={Journal of Applied Physics},
  volume={138},
  number={11},
  year={2025},
  publisher={AIP Publishing}
}

@article{faysal2026first,
  title={First-principles study of Y2CCl2 and Janus Y2CClX (X= F, Br, I) MXenes for photovoltaic applications},
  author={Faysal, AA and Ghosh, Swarup and Lee, It Ee and Wali, Qamar and Aamir, Muhammad and Ali, MA and Hossain, MM and Jahan, N and Hossain, MZ and Akhtaruzzaman, Md and others},
  journal={Scientific Reports},
  year={2026},
  publisher={Nature Publishing Group UK, London}
}

@article{akgenc2021tuning,
  title={Tuning of electronic structure, magnetic phase, and transition temperature in two-dimensional Cr-based Janus MXenes},
  author={Akgenc, B and Vatansever, EROL and Ersan, F},
  journal={Physical Review Materials},
  volume={5},
  number={8},
  pages={083403},
  year={2021},
  publisher={APS}
}

@article{he2021tb2j,
  title={TB2J: A python package for computing magnetic interaction parameters},
  author={He, Xu and Helbig, Nicole and Verstraete, Matthieu J and Bousquet, Eric},
  journal={Computer Physics Communications},
  volume={264},
  pages={107938},
  year={2021},
  publisher={Elsevier}
}

@article{rezaei2026benchmarking,
  title={Benchmarking first-principles approaches for extracting magnetic exchange interactions},
  author={Rezaei, Nafise and Oganov, Artem R and Ghojavand, Ali and Milo{\v{s}}evi{\'c}, Milorad V and Alaei, Mojtaba},
  journal={npj Computational Materials},
  year={2026},
  publisher={Nature Publishing Group}
}

@article{evans2014atomistic,
  title={Atomistic spin model simulations of magnetic nanomaterials},
  author={Evans, Richard FL and Fan, Weijia J and Chureemart, Phanwadee and Ostler, Thomas A and Ellis, Matthew OA and Chantrell, Roy W},
  journal={Journal of Physics: Condensed Matter},
  volume={26},
  number={10},
  pages={103202},
  year={2014},
  publisher={IOP Publishing}
}

@article{barik2026magnetic,
  title={Magnetic properties of the J eff= 1 2 based frustrated triangular lattice antiferromagnet Ba 4 YbReWO 12},
  author={Barik, M and Khatua, J and Kim, Suyoung and Mun, Eundeok and Lee, Suheon and Hitti, Bassam and Morris, Gerald D and Choi, Kwang-Yong and Khuntia, P},
  journal={Physical Review B},
  volume={113},
  number={2},
  pages={024434},
  year={2026},
  publisher={APS}
}

@article{starykh2015unusual,
  title={Unusual ordered phases of highly frustrated magnets: a review},
  author={Starykh, Oleg A},
  journal={Reports on Progress in Physics},
  volume={78},
  number={5},
  pages={052502},
  year={2015},
  publisher={IOP Publishing}
}

@article{wang2025intriguing,
  title={Intriguing kagome topological materials},
  author={Wang, Qi and Lei, Hechang and Qi, Yanpeng and Felser, Claudia},
  journal={npj Quantum Materials},
  volume={10},
  number={1},
  pages={72},
  year={2025},
  publisher={Nature Publishing Group UK, London}
}

@article{jungwirth2016antiferromagnetic,
  title={Antiferromagnetic spintronics},
  author={Jungwirth, Tomas and Marti, Xavier and Wadley, Patrick and Wunderlich, Jairo},
  journal={Nature nanotechnology},
  volume={11},
  number={3},
  pages={231--241},
  year={2016},
  publisher={Nature Publishing Group}
}

@article{aldarawsheh2022emergence,
  title={Emergence of zero-field non-synthetic single and interchained antiferromagnetic skyrmions in thin films},
  author={Aldarawsheh, Amal and Fernandes, Imara Lima and Brinker, Sascha and Sallermann, Moritz and Abusaa, Muayad and Bl{\"u}gel, Stefan and Lounis, Samir},
  journal={Nature Communications},
  volume={13},
  number={1},
  pages={7369},
  year={2022},
  publisher={Nature Publishing Group UK, London}
}

@article{nvemec2018antiferromagnetic,
  title={Antiferromagnetic opto-spintronics},
  author={N{\v{e}}mec, Petr and Fiebig, Manfred and Kampfrath, Tobias and Kimel, Alexey V},
  journal={Nature Physics},
  volume={14},
  number={3},
  pages={229--241},
  year={2018},
  publisher={Nature Publishing Group UK, London}
}

@article{gomonay2017concepts,
  title={Concepts of antiferromagnetic spintronics},
  author={Gomonay, Olena and Jungwirth, T and Sinova, Jairo},
  journal={physica status solidi (RRL)--Rapid Research Letters},
  volume={11},
  number={4},
  pages={1700022},
  year={2017},
  publisher={Wiley Online Library}
}

@article{gomonay2018antiferromagnetic,
  title={Antiferromagnetic spin textures and dynamics},
  author={Gomonay, O and Baltz, V and Brataas, A and Tserkovnyak, Y},
  journal={Nature Physics},
  volume={14},
  number={3},
  pages={213--216},
  year={2018},
  publisher={Nature Publishing Group UK, London}
}

@article{kampfrath2011coherent,
  title={Coherent terahertz control of antiferromagnetic spin waves},
  author={Kampfrath, Tobias and Sell, Alexander and Klatt, Gregor and Pashkin, Alexej and M{\"a}hrlein, Sebastian and Dekorsy, Thomas and Wolf, Martin and Fiebig, Manfred and Leitenstorfer, Alfred and Huber, Rupert},
  journal={Nature Photonics},
  volume={5},
  number={1},
  pages={31--34},
  year={2011},
  publisher={Nature Publishing Group UK, London}
}

@article{baierl2016terahertz,
  title={Terahertz-driven nonlinear spin response of antiferromagnetic nickel oxide},
  author={Baierl, Sebastian and Mentink, Johan H and Hohenleutner, Matthias and Braun, Lukas and Do, T-M and Lange, Christoph and Sell, Alexander and Fiebig, Manfred and Woltersdorf, Georg and Kampfrath, Tobias and others},
  journal={Physical review letters},
  volume={117},
  number={19},
  pages={197201},
  year={2016},
  publisher={APS}
}

@article{bhattacharjee2018neel,
  title={N{\'e}el spin-orbit torque driven antiferromagnetic resonance in Mn 2 Au probed by time-domain THz spectroscopy},
  author={Bhattacharjee, N and Sapozhnik, AA and Bodnar, S Yu and Grigorev, V Yu and Agustsson, SY and Cao, J and Dominko, D and Obergfell, M and Gomonay, O and Sinova, J and others},
  journal={Physical review letters},
  volume={120},
  number={23},
  pages={237201},
  year={2018},
  publisher={APS}
}

@article{legrand2020room,
  title={Room-temperature stabilization of antiferromagnetic skyrmions in synthetic antiferromagnets},
  author={Legrand, William and Maccariello, Davide and Ajejas, Fernando and Collin, Sophie and Vecchiola, Aymeric and Bouzehouane, Karim and Reyren, Nicolas and Cros, Vincent and Fert, Albert},
  journal={Nature materials},
  volume={19},
  number={1},
  pages={34--42},
  year={2020},
  publisher={Nature Publishing Group UK, London}
}

@article{barker2016static,
  title={Static and dynamical properties of antiferromagnetic skyrmions in the presence of applied current and temperature},
  author={Barker, Joseph and Tretiakov, Oleg A},
  journal={Physical review letters},
  volume={116},
  number={14},
  pages={147203},
  year={2016},
  publisher={APS}
}

@article{juge2022skyrmions,
  title={Skyrmions in synthetic antiferromagnets and their nucleation via electrical current and ultra-fast laser illumination},
  author={Juge, Rom{\'e}o and Sisodia, Naveen and Larra{\~n}aga, Joseba Urrestarazu and Zhang, Qiang and Pham, Van Tuong and Rana, Kumari Gaurav and Sarpi, Brice and Mille, Nicolas and Stanescu, Stefan and Belkhou, Rachid and others},
  journal={Nature Communications},
  volume={13},
  number={1},
  pages={4807},
  year={2022},
  publisher={Nature Publishing Group UK, London}
}

@article{rosales2015three,
  title={Three-sublattice skyrmion crystal in the antiferromagnetic triangular lattice},
  author={Rosales, H{\'e}ctor Diego and Cabra, Daniel Carlos and Pujol, Pierre},
  journal={Physical Review B},
  volume={92},
  number={21},
  pages={214439},
  year={2015},
  publisher={APS}
}

@article{zhang2016antiferromagnetic,
  title={Antiferromagnetic skyrmion: stability, creation and manipulation},
  author={Zhang, Xichao and Zhou, Yan and Ezawa, Motohiko},
  journal={Scientific reports},
  volume={6},
  number={1},
  pages={24795},
  year={2016},
  publisher={Nature Publishing Group UK, London}
}

@article{jin2016dynamics,
  title={Dynamics of antiferromagnetic skyrmion driven by the spin Hall effect},
  author={Jin, Chendong and Song, Chengkun and Wang, Jianbo and Liu, Qingfang},
  journal={Applied Physics Letters},
  volume={109},
  number={18},
  year={2016},
  publisher={AIP Publishing}
}

@article{fernandes2019skyrmions,
  title={Skyrmions and merons in two-dimensional antiferromagnetic systems},
  author={Fernandes, RL and Lopes, RJC and Pereira, AR},
  journal={Solid State Communications},
  volume={290},
  pages={55--59},
  year={2019},
  publisher={Elsevier}
}

@article{shen2020current,
  title={Current-induced dynamics and chaos of antiferromagnetic bimerons},
  author={Shen, Laichuan and Xia, Jing and Zhang, Xichao and Ezawa, Motohiko and Tretiakov, Oleg A and Liu, Xiaoxi and Zhao, Guoping and Zhou, Yan},
  journal={Physical review letters},
  volume={124},
  number={3},
  pages={037202},
  year={2020},
  publisher={APS}
}

@article{silva2021antiferromagnetic,
  title={Antiferromagnetic-Bimeron dynamics driven by a spin-polarized current at an inhomogeneous racetrack},
  author={Silva, RL},
  journal={Physics Letters A},
  volume={403},
  pages={127399},
  year={2021},
  publisher={Elsevier}
}

@article{akosa2017theory,
  title={Theory of topological spin Hall effect in antiferromagnetic skyrmion: impact on current-induced motion},
  author={Akosa, Collins Ashu and Tretiakov, OA and Tatara, G and Manchon, Aurelien},
  journal={arXiv preprint arXiv:1709.02931},
  year={2017}
}

@article{litzius2017skyrmion,
  title={Skyrmion Hall effect revealed by direct time-resolved X-ray microscopy},
  author={Litzius, Kai and Lemesh, Ivan and Kr{\"u}ger, Benjamin and Bassirian, Pedram and Caretta, Lucas and Richter, Kornel and B{\"u}ttner, Felix and Sato, Koji and Tretiakov, Oleg A and F{\"o}rster, Johannes and others},
  journal={Nature Physics},
  volume={13},
  number={2},
  pages={170--175},
  year={2017},
  publisher={Nature Publishing Group UK, London}
}

@article{lin2013particle,
  title={Particle model for skyrmions in metallic chiral magnets: Dynamics, pinning, and creep},
  author={Lin, Shi-Zeng and Reichhardt, Charles and Batista, Cristian D and Saxena, Avadh},
  journal={Physical Review B—Condensed Matter and Materials Physics},
  volume={87},
  number={21},
  pages={214419},
  year={2013},
  publisher={APS}
}

@article{dohi2019formation,
  title={Formation and current-induced motion of synthetic antiferromagnetic skyrmion bubbles},
  author={Dohi, Takaaki and DuttaGupta, Samik and Fukami, Shunsuke and Ohno, Hideo},
  journal={Nature communications},
  volume={10},
  number={1},
  pages={5153},
  year={2019},
  publisher={Nature Publishing Group UK, London}
}

@article{chen2022controllable,
  title={Controllable generation of antiferromagnetic skyrmions in synthetic antiferromagnets with thermal effect},
  author={Chen, Ruyi and Cui, Qirui and Han, Lei and Xue, Xiaotian and Liang, Jinghua and Bai, Hua and Zhou, Yongjian and Pan, Feng and Yang, Hongxin and Song, Cheng},
  journal={Advanced Functional Materials},
  volume={32},
  number={17},
  pages={2111906},
  year={2022},
  publisher={Wiley Online Library}
}

@article{zhang2023giant,
  title={Giant Dzyaloshinskii-Moriya interaction, strong XXZ-type biquadratic coupling, and bimeronic excitations in the two-dimensional CrMnI6 magnet},
  author={Zhang, Shunhong and Li, Xiaoyin and Zhang, Huisheng and Cui, Ping and Xu, Xiaohong and Zhang, Zhenyu},
  journal={npj Quantum Materials},
  volume={8},
  number={1},
  pages={38},
  year={2023},
  publisher={Nature Publishing Group UK, London}
}

@article{muller2019spirit,
  title={Spirit: Multifunctional framework for atomistic spin simulations},
  author={M{\"u}ller, Gideon P and Hoffmann, Markus and Di{\ss}elkamp, Constantin and Sch{\"u}rhoff, Daniel and Mavros, Stefanos and Sallermann, Moritz and Kiselev, Nikolai S and J{\'o}nsson, Hannes and Bl{\"u}gel, Stefan},
  journal={Physical review b},
  volume={99},
  number={22},
  pages={224414},
  year={2019},
  publisher={APS}
}

@article{koshibae2016berry,
  title={Berry curvature and dynamics of a magnetic bubble},
  author={Koshibae, Wataru and Nagaosa, Naoto},
  journal={New Journal of Physics},
  volume={18},
  number={4},
  pages={045007},
  year={2016},
  publisher={IOP Publishing}
}

@article{zhou2024kagomerization,
  title={Kagomerization of transition metal monolayers induced by two-dimensional hexagonal boron nitride},
  author={Zhou, Hangyu and dos Santos Dias, Manuel and Zhang, Youguang and Zhao, Weisheng and Lounis, Samir},
  journal={nature communications},
  volume={15},
  number={1},
  pages={4854},
  year={2024},
  publisher={Nature Publishing Group UK, London}
}

@article{abuawwad2026chiral,
  title={Chiral spin-textures in van der Waals heterostructures},
  author={AbuAwwad, Nihad and Lounis, Samir},
  journal={2D Materials},
  year={2026}
}

@article{abuawwad2023crte,
  title={CrTe 2 as a two-dimensional material for topological magnetism in complex heterobilayers},
  author={Abuawwad, Nihad and Dias, Manuel dos Santos and Abusara, Hazem and Lounis, Samir},
  journal={Physical Review B},
  volume={108},
  number={9},
  pages={094409},
  year={2023},
  publisher={APS}
}

@article{wysokinski2019mechanism,
  title={Mechanism for transitions between ferromagnetic and antiferromagnetic orders in d-electron metallic magnets},
  author={Wysoki{\'n}ski, Marcin M},
  journal={Scientific reports},
  volume={9},
  number={1},
  pages={19461},
  year={2019},
  publisher={Nature Publishing Group UK, London}
}

@article{dugu2026interplay,
  title={Interplay of ferromagnetic and antiferromagnetic interactions in epitaxial Co3ZnN},
  author={Dugu, Sita and Mahatara, Sharad and Regier, Corlyn E and Neilson, James R and Lany, Stephan and Smaha, Rebecca W and Bauers, Sage R},
  journal={Applied Physics Letters},
  volume={128},
  number={20},
  year={2026},
  publisher={AIP Publishing}
}

@article{friedemann2018quantum,
  title={Quantum tricritical points in NbFe2},
  author={Friedemann, Sven and Duncan, Will J and Hirschberger, Max and Bauer, Thomas W and K{\"u}chler, Robert and Neubauer, Andreas and Brando, Manuel and Pfleiderer, Christian and Grosche, F Malte},
  journal={Nature Physics},
  volume={14},
  number={1},
  pages={62--67},
  year={2018},
  publisher={Nature Publishing Group UK, London}
}

@article{taufour2016ferromagnetic,
  title={Ferromagnetic quantum critical point avoided by the appearance of another magnetic phase in LaCrGe 3 under pressure},
  author={Taufour, Valentin and Kaluarachchi, Udhara S and Khasanov, Rustem and Nguyen, Manh Cuong and Guguchia, Zurab and Biswas, Pabitra Kumar and Bonf{\`a}, Pietro and De Renzi, Roberto and Lin, Xiao and Kim, Stella K and others},
  journal={Physical review letters},
  volume={117},
  number={3},
  pages={037207},
  year={2016},
  publisher={APS}
}

@article{kaluarachchi2017tricritical,
  title={Tricritical wings and modulated magnetic phases in LaCrGe3 under pressure},
  author={Kaluarachchi, Udhara S and Bud’ko, Sergey L and Canfield, Paul C and Taufour, Valentin},
  journal={Nature communications},
  volume={8},
  number={1},
  pages={546},
  year={2017},
  publisher={Nature Publishing Group UK, London}
}

@article{jeffries2016emergent,
  title={Emergent ferromagnetism and T-linear scattering in USb 2 at high pressure},
  author={Jeffries, Jason R and Stillwell, Ryan L and Weir, Samuel T and Vohra, Yogesh K and Butch, Nicholas P},
  journal={Physical Review B},
  volume={93},
  number={18},
  pages={184406},
  year={2016},
  publisher={APS}
}

@article{pospivsil2017switching,
  title={Switching of magnetic ground states across the UI r 1- x R hx Ge alloy system},
  author={Posp{\'\i}{\v{s}}il, Ji{\v{r}}{\'\i} and Haga, Yoshinori and Kambe, Shinsaku and Tokunaga, Yo and Tateiwa, Naoyuki and Aoki, Dai and Honda, Fuminori and Nakamura, Ai and Homma, Yoshiya and Yamamoto, Etsuji and others},
  journal={Physical Review B},
  volume={95},
  number={15},
  pages={155138},
  year={2017},
  publisher={APS}
}

@article{lengyel2015avoided,
  title={Avoided ferromagnetic quantum critical point in CeRuPO},
  author={Lengyel, E and Macovei, ME and Jesche, Anton and Krellner, C and Geibel, Christoph and Nicklas, Michael},
  journal={Physical Review B},
  volume={91},
  number={3},
  pages={035130},
  year={2015},
  publisher={APS}
}

@article{rybakov2019chiral,
  title={Chiral magnetic skyrmions with arbitrary topological charge},
  author={Rybakov, Filipp N and Kiselev, Nikolai S},
  journal={Physical review B},
  volume={99},
  number={6},
  pages={064437},
  year={2019},
  publisher={APS}
}

@article{nunez2006theory,
  title={Theory of spin torques and giant magnetoresistance in antiferromagnetic metals},
  author={N{\'u}nez, Alvaro S and Duine, RA and Haney, Paul and MacDonald, AH},
  journal={Physical Review B—Condensed Matter and Materials Physics},
  volume={73},
  number={21},
  pages={214426},
  year={2006},
  publisher={APS}
}

@article{gomonay2012symmetry,
  title={Symmetry and the macroscopic dynamics of antiferromagnetic materials in the presence of spin-polarized current},
  author={Gomonay, Helen V and Kunitsyn, Roman V and Loktev, Vadim M},
  journal={Physical Review B—Condensed Matter and Materials Physics},
  volume={85},
  number={13},
  pages={134446},
  year={2012},
  publisher={APS}
}

@article{dal2024antiferromagnetic,
  title={Antiferromagnetic spintronics and beyond},
  author={Dal Din, A and Amin, OJ and Wadley, P and Edmonds, KW},
  journal={npj Spintronics},
  volume={2},
  number={1},
  pages={25},
  year={2024},
  publisher={Nature Publishing Group UK, London}
}

@article{amrillah2023potential,
  title={Potential of MXenes as a novel material for spintronic devices: a review},
  author={Amrillah, Tahta and Hermawan, Angga and Cristian, Yeremia Budi and Oktafiani, Agustina and Dewi, Diva Meisya Maulina and Amalina, Ilma and Darminto and Juang, Jenh-Yih},
  journal={Physical Chemistry Chemical Physics},
  volume={25},
  number={28},
  pages={18584--18608},
  year={2023},
  publisher={The Royal Society of Chemistry}
}

@article{kim2026magnetic,
  title={Magnetic MXene-based microrobots for soil microplastics removal},
  author={Kim, Jeonghyo and Mallick, Apabrita and Song, Su-Jin and Pumera, Martin},
  journal={NPG Asia Materials},
  year={2026},
  publisher={Springer Japan, Tokyo}
}

@article{garcia2023theoretical,
  title={Theoretical analysis of magnetic coupling in the Ti2C bare MXene},
  author={Garcia-Romeral, Nestor and Morales-Garcia, Angel and Vines, Francesc and Moreira, Iberio de PR and Illas, Francesc},
  journal={The Journal of Physical Chemistry C},
  volume={127},
  number={7},
  pages={3706--3714},
  year={2023},
  publisher={ACS Publications}
}

@article{chen2023two,
  title={Two-dimensional janus MXene inks for versatile functional coatings on arbitrary substrates},
  author={Chen, Mengjie and Li, Lulu and Deng, Zhiming and Min, Peng and Yu, Zhong-Zhen and Zhang, Chuanfang John and Zhang, Hao-Bin},
  journal={ACS Applied Materials \& Interfaces},
  volume={15},
  number={3},
  pages={4591--4600},
  year={2023},
  publisher={ACS Publications}
}

@article{qian2025biomimetic,
  title={Biomimetic Janus MXene membrane with bidirectional ion permselectivity for enhanced osmotic effects and iontronic logic control},
  author={Qian, Han and Fan, Hongzhao and Peng, Puguang and Du, Yan and Li, Xiang and Liu, Yanhui and Yang, Feiyao and Zhou, Yanguang and Wang, Zhong Lin and Wei, Di},
  journal={Science Advances},
  volume={11},
  number={38},
  pages={eadx1184},
  year={2025},
  publisher={American Association for the Advancement of Science}
}

@article{ram2024tuning,
  title={Tuning the electronic and magnetic properties of double transition metal (MCrCT2, M= Ti, Mo) Janus MXenes for enhanced spintronics and nanoelectronics},
  author={Ram, Swetarekha and Koshi, Namitha Anna and Lee, Seung-Cheol and Bhattacharjee, Satadeep},
  journal={The Journal of Physical Chemistry C},
  volume={128},
  number={17},
  pages={7323--7333},
  year={2024},
  publisher={ACS Publications}
}

@article{limbu2025magnetic,
  title={Magnetic semiconductors from functionalized Cr 2 C MXenes},
  author={Limbu, Yogendra and Paudyal, Hari and da Silva, Eudes Gomes and Candido, Denis R and Flatt{\'e}, Michael E and Paudyal, Durga},
  journal={Physical Review B},
  volume={112},
  number={21},
  pages={214445},
  year={2025},
  publisher={APS}
}

@article{PhysRevB.73.045112,
  title = {Linear optical properties in the projector-augmented wave methodology},
  author = {Gajdo{\ifmmode \check{s}\else \v{s}\fi{}}, M. and Hummer, K. and Kresse, G. and Furthm{\"u}ller, J. and Bechstedt, F.},
  journal = {Phys. Rev. B},
  volume = {73},
  issue = {4},
  pages = {045112},
  numpages = {9},
  year = {2006},
  month = {Jan},
  publisher = {American Physical Society},
  doi = {10.1103/PhysRevB.73.045112},
  url = {https://link.aps.org/doi/10.1103/PhysRevB.73.045112}
}

@article{matsumoto2011theoretical,
  title={Theoretical prediction of a rotating magnon wave packet in ferromagnets},
  author={Matsumoto, Ryo and Murakami, Shuichi},
  journal={Physical review letters},
  volume={106},
  number={19},
  pages={197202},
  year={2011},
  publisher={APS}
}

@article{murakami2017thermal,
  title={Thermal Hall effect of magnons},
  author={Murakami, Shuichi and Okamoto, Akihiro},
  journal={Journal of the Physical Society of Japan},
  volume={86},
  number={1},
  pages={011010},
  year={2017},
  publisher={The Physical Society of Japan}
}

@article{colpa1978diagonalization,
  title={Diagonalization of the quadratic boson hamiltonian},
  author={Colpa, JHP},
  journal={Physica A: Statistical Mechanics and its Applications},
  volume={93},
  number={3-4},
  pages={327--353},
  year={1978},
  publisher={Elsevier}
}

@article{shindou2013topological,
  title={Topological chiral magnonic edge mode in a magnonic crystal},
  author={Shindou, Ryuichi and Matsumoto, Ryo and Murakami, Shuichi and Ohe, Jun-ichiro},
  journal={Physical Review B—Condensed Matter and Materials Physics},
  volume={87},
  number={17},
  pages={174427},
  year={2013},
  publisher={APS}
}

@article{fukui2005chern,
  title={Chern numbers in discretized Brillouin zone: Efficient method of computing (spin) Hall conductances},
  author={Fukui, Takahiro and Hatsugai, Yasuhiro and Suzuki, Hiroshi},
  journal={Journal of the Physical Society of Japan},
  volume={74},
  number={6},
  pages={1674--1677},
  year={2005},
  publisher={The Physical Society of Japan}
}

@article{holstein1940field,
  title={Field dependence of the intrinsic domain magnetization of a ferromagnet},
  author={Holstein, Theodore and Primakoff, Henry},
  journal={Physical Review},
  volume={58},
  number={12},
  pages={1098},
  year={1940},
  publisher={APS}
}

@article{rybakov2026quantum,
  title={Quantum Spin Wave Theory with Multi-Spin Interactions for Magnetic Materials Modeling},
  author={Rybakov, Andrey and Marino, Marco and Blanter, Yaroslav M and Coronado, Eugenio and Ferrer, Jaime},
  journal={arXiv preprint arXiv:2611.06833},
  year={2026}
}

@article{bhukta2026time,
  title={Time-resolved imaging of antiferromagnetic skyrmion interactions},
  author={Bhukta, Mona and Dohi, Takaaki and Leutner, Kilian and Syskaki, Maria-Andromachi and Kammerbauer, Fabian and Tran, Duc Minh and Wintz, Sebastian and Weigand, Markus and Finizio, Simone and Raabe, J{\"o}rg and others},
  journal={Nature Physics},
  pages={1--9},
  year={2026},
  publisher={Nature Publishing Group UK, London}
}

@article{pan2024strain,
  title={Strain-induced intrinsic antiferromagnetic skyrmions in two-dimensional Janus magnets},
  author={Pan, Weiyi and Ji, Shilei and Xu, Zhiming},
  journal={Physical Review B},
  volume={110},
  number={14},
  pages={144408},
  year={2024},
  publisher={APS}
}

@article{jani2021antiferromagnetic,
  title={Antiferromagnetic half-skyrmions and bimerons at room temperature},
  author={Jani, Hariom and Lin, Jheng-Cyuan and Chen, Jiahao and Harrison, Jack and Maccherozzi, Francesco and Schad, Jonathon and Prakash, Saurav and Eom, Chang-Beom and Ariando, Ariando and Venkatesan, Thirumalai and others},
  journal={Nature},
  volume={590},
  number={7844},
  pages={74--79},
  year={2021},
  publisher={Nature Publishing Group UK, London}
}

@article{li2025intrinsic,
  title={Intrinsic antiferromagnetic skyrmions and their strain modulation in atomically thin Janus vanadium dihalides},
  author={Li, Chuyu and Zhou, Wenzhe and Li, Yating and He, Xianjuan and Chen, Yahan and Ouyang, Fangping},
  journal={Physical Review B},
  volume={112},
  number={1},
  pages={014424},
  year={2025},
  publisher={APS}
}

@article{romming2013writing,
  title={Writing and deleting single magnetic skyrmions},
  author={Romming, Niklas and Hanneken, Christian and Menzel, Matthias and Bickel, Jessica E and Wolter, Boris and Von Bergmann, Kirsten and Kubetzka, Andr{\'e} and Wiesendanger, Roland},
  journal={Science},
  volume={341},
  number={6146},
  pages={636--639},
  year={2013},
  publisher={American Association for the Advancement of Science}
}

@article{spethmann2020discovery,
  title={Discovery of magnetic single-and triple-q states in Mn/Re (0001)},
  author={Spethmann, Jonas and Meyer, Sebastian and von Bergmann, Kirsten and Wiesendanger, Roland and Heinze, Stefan and Kubetzka, Andr{\'e}},
  journal={Physical review letters},
  volume={124},
  number={22},
  pages={227203},
  year={2020},
  publisher={APS}
}

\end{document}